\documentclass[
  journal=pasa,
  manuscript=research-paper, 
  year=2023,
  volume=00,
]{cup-journal}

\usepackage{microtype,siunitx,booktabs}
\usepackage[table,xcdraw]{}
\usepackage{IEEEtrantools}
\usepackage[colorlinks=true,linkcolor=blue, citecolor=blue, urlcolor=blue]{hyperref}
\usepackage{color,soul}
\usepackage{multirow}
\usepackage{dirtree}
\usepackage{comment}
\usepackage{amsmath}
\usepackage{threeparttable}
\usepackage{subfig}
\usepackage{makecell}
\usepackage[nolist]{acronym}
\usepackage[export]{adjustbox}
\usepackage[
singlelinecheck=false 
]{caption}
\usepackage{flushend}

\DeclareRobustCommand{\ion}[2]{%
\relax\ifmmode
\ifx\testbx\f@series
{\mathbf{#1\,\mathsc{#2}}}\else
{\mathrm{#1\,\mathsc{#2}}}\fi
\else\textup{#1\,{\mdseries\textsc{#2}}}%
\fi}

\title{Survey and Monitoring of ASKAP's RFI Environment and Trends I: Flagging Statistics}

\author{L. Louren\c{c}o}
\affiliation{Sydney Institute for Astronomy, School of Physics, University of Sydney, Sydney, New South Wales 2006, Australia.}
\alsoaffiliation{CSIRO Space and Astronomy, PO Box 76, Epping, New South Wales 1710, Australia}
\email[L. Louren\c{c}o]{liroy.lourenco@sydney.edu.au, liroy.lourenco@csiro.au}

\author{A.~P. Chippendale}
\affiliation{CSIRO Space and Astronomy, PO Box 76, Epping, New South Wales 1710, Australia}

\author{B. Indermuehle}
\affiliation{CSIRO Space and Astronomy, PO Box 76, Epping, New South Wales 1710, Australia}

\author{V.~A. Moss}
\affiliation{CSIRO Space and Astronomy, PO Box 76, Epping, New South Wales 1710, Australia}

\author{Tara Murphy}
\affiliation{Sydney Institute for Astronomy, School of Physics, University of Sydney, Sydney, New South Wales 2006, Australia.}

\author{T.~J. Galvin}
\affiliation{ATNF, CSIRO Space \& Astronomy, PO Box 1130, Bentley, WA 6102, Australia}

\author{G. Hellbourg}
\affiliation{Cahill Center for Astronomy and Astrophysics, MC 249-17 California Institute of Technology, Pasadena, CA 91125, USA}

\author{A.~W. Hotan}
\affiliation{CSIRO Space and Astronomy, PO Box 76, Epping, New South Wales 1710, Australia}

\author{E. Lenc}
\affiliation{CSIRO Space and Astronomy, PO Box 76, Epping, New South Wales 1710, Australia}

\author{M.~T. Whiting}
\affiliation{CSIRO Space and Astronomy, PO Box 76, Epping, New South Wales 1710, Australia}

\doi{10.1017/pasa.2023.xxx}

\received {dd Mmm YYYY}
\revised  {dd Mmm YYYY}
\accepted {dd Mmm YYYY}
\published{22 September 2023}


\newcommand{\nhours}{\qty{1500}{\hour}~}
\newcommand{\nSBIDs}{\num{5000}~}
\newcommand{\dates}{April 2019 to August 2023}
\newcommand{\percentFlagged}{\qty{3}{\percent}~}

\begin{document}
\begin{acronym}
    \acro{ASKAP}{Australian SKA Pathfinder}
    \acro{RFI}{Radio Frequency Interference}
    \acro{CSIRO}{Commonwealth Scientific and Industrial Research Organisation}
    \acro{MRO}{Murchison Radio-astronomy Observatory}
    \acro{SMART}{``Survey and Monitoring of \ac{ASKAP}'s \ac{RFI} environment and Trends''}
    \acro{ARQZWA}{Australian Radio Quiet Zone – Western Australia}
    \acro{ITU}{International Telecommunication Union}
    \acro{ACMA}{Australian Communications and Media Authority}
    \acro{POSSUM}{POlarisation Sky Survey of the Universe’s Magnetism}
    \acro{RACS}{Rapid \ac{ASKAP} Continuum Survey}
    \acro{VAST}{\ac{ASKAP} survey for Variables and Slow Transients}
    \acro{EMU}{Evolutionary Map of the Universe}
    \acro{CASDA}{\acs{CSIRO} \ac{ASKAP} Science Data Archive}
    \acro{MWA}{Murchison Widefield Array}
    \acro{FAST}{Five-hundred-meter Aperture Spherical radio Telescope}
    \acro{GUI}{Graphical User Interface}
    \acro{DME}{Distance Measuring Equipment}
    \acro{GPS}{Global Positioning Satellite} 
    \acro{ADS-B}{Automatic Dependent Surveillance–Broadcast}
    \acro{GLONASS}{GLObalnaya NAvigatsionnaya Sputnikovaya Sistema in Russian}
    \acro{UTC}{Coordinated Universal Time}
    \acro{AWST}{Australian Western Standard Time}
    \acro{PAF}{Phased Array Feed}
    \acro{SSP}{Survey Science Project}
    \acro{RAS}{radio astronomy service}
    \acro{RR}{radio regulation}
    \acro{SAURON}{Scheduling Autonomously Under Reactive Observational Needs}
    \acro{ATNF}{Australia Telescope National Facility}
    \acro{FPGA}{Field Programmable Gate Array}
    \acro{ODC}{On-Dish Calibrator}
    \acro{REMP}{Radio Emissions Management Plan}
    \acro{RMS}{Root Mean Square}
    \acro{LEO}{Low Earth Orbit}
    \acro{UEMR}{unintended electromagnetic radiation}
\end{acronym}

\begin{abstract}
We present an initial analysis of \ac{RFI} flagging statistics from archived \ac{ASKAP} observations for the \ac{SMART} project. \ac{SMART} is a two-part observatory-led project combining analysis of archived observations with a dedicated, comprehensive \ac{RFI} survey. The survey component covers \ac{ASKAP}'s full \qtyrange{700}{1800}{\mega\hertz} frequency range, including bands not typically used due to severe \ac{RFI}. Observations are underway to capture a detailed snapshot of the \ac{ASKAP} \ac{RFI} environment over representative \qty{24}{\hour} periods. In addition to this dedicated survey, we routinely archive and analyse flagging statistics for all scientific observations to monitor the observatory's \ac{RFI} environment in near real-time.  We use the telescope itself as a very sensitive \ac{RFI} monitor and directly assess the fraction of scientific observations impacted by \ac{RFI}. To this end, flag tables are now automatically ingested and aggregated as part of routine \ac{ASKAP} operations for all science observations, as a function of frequency and time.
The data presented in this paper come from processing all archived data for several \ac{ASKAP} \acp{SSP}. We found that the average amount of flagging due to \ac{RFI} across the routinely-used `clean' continuum science bands is \percentFlagged. The `clean' mid band from \qtyrange{1293}{1437}{\mega\hertz} (excluding the \qty{144}{\mega\hertz} below \qty{1293}{\mega\hertz} impacted by radionavigation-satellites which is discarded before processing) is the least affected by \ac{RFI}, followed by the `clean' low band from \qtyrange{742}{1085}{\mega\hertz}. \ac{ASKAP} \acp{SSP} lose most of their data to the mobile service in the low band, aeronautical service in the mid band and satellite navigation service in the \qtyrange{1510}{1797}{\mega\hertz} high band. We also show that for some of these services, the percentage of discarded data has been increasing year-on-year. 

\ac{SMART} provides a unique opportunity to study \ac{ASKAP}'s changing \ac{RFI} environment, including understanding and updating the default flagging behaviour, inferring the suitability of and calibrating \ac{RFI} monitoring equipment, monitoring spectrum management compliance in the \ac{ARQZWA}, and informing the implementation of a suite of \ac{RFI} mitigation techniques.
\end{abstract}
 
\section{Introduction}
\label{sec:int}

\acresetall
Powerful human-generated transmissions corrupt weaker radio signals from cosmic sources. Balancing the advantages of wireless communication, satellite navigation, consumer electronics, weather forecasting and air transport with novel techniques \citep{Zheleva2023RadioCoexistence} to manage \ac{RFI} is essential in keeping scientific research in radio astronomy sustainable in the future. The challenge for radio astronomy is not only the comparative power of these unwanted signals but also their increasing prevalence \citep{CommitteeonRadioAstronomyFrequencies1997CRAFAstronomy}. From a scientific perspective, all undesired signals are described as \ac{RFI}. However, practically (and legislatively) emissions are only classified as \ac{RFI} if they impinge on the authorised user of  bands allocated (or shared) amongst different services (e.g. the \ac{RAS}, aeronautical radionavigation service, broadcasting service, fixed/mobile  service, etc.). In Australia this is under the jurisdiction of the \ac{ACMA}, and of the \ac{ITU} treaty, implemented by way of the \acp{RR} \citep{Baan2019ImplementingScience, RR2016}. To further complicate the issue some \ac{RFI} is a result of \acl{UEMR}, emission outside the service allocation of the intended transmission and therefore not subject to the same regulations stipulated above \citep{DiVrunoF.2023UnintendedMHz, GriggD.2023DetectionAnalogue}.

There are many methods of \ac{RFI} mitigation  \citep{Fridman2001RFIAstronomy, Kesteven2010OverviewSystems, Series2013TechniquesAstronomy, Baan2019ImplementingScience} depending on the type of \ac{RFI} and instrument. Indeed, observatories have to use a variety of techniques at different stages in the signal chain to manage \ac{RFI} \citep{Baan2019ImplementingScience}. Flagging is the most common strategy to mitigate the effects of \ac{RFI} in radio astronomy; it is the process of identifying and discarding corrupt data in the frequency or time domain. Flagging can be conducted manually but is increasingly done by more sophisticated flagging software \citep{Offringa2010Post-correlationMethods, Burd2018DetectingAlgorithm}. Unfortunately, flagging reduces measurement sensitivity and the discarded data may contain valuable information from astronomical sources.  Flagging also impacts the instrumental response in ways that may require consideration when drawing scientific conclusions.  An example is that flagging impacts the UV coverage and thereby the point spread function of a synthesis imaging array like the \ac{ASKAP}.  Furthermore, newer generations of radio telescopes with increased sensitivity, larger bandwidths and wider fields of view --- designed to keep up with contemporary science goals --- are more severely affected by \ac{RFI}, resulting in even more flagging. 

The \ac{ASKAP} telescope, operated by the \acs{CSIRO}, is one such instrument \citep{Hotan2021AustralianDescription}. ASKAP operates over the \qtyrange{700}{1800}{\mega\hertz} band of which less than \qty{4}{\percent} is allocated to the radio astronomy service (RAS) on a primary basis in three small protected bands:
\begin{enumerate}
\item \qtyrange{1400}{1427}{\mega\hertz} (Hydrogen line), 
\item \qtyrange{1610.6}{1613.8}{\mega\hertz} (OH line), and 
\item \qtyrange{1660}{1670}{\mega\hertz} (OH line). 
\end{enumerate} 
The remainder of the frequencies \ac{ASKAP} operates are in use by the fixed/mobile service, the radio location service, the aeronautical mobile service, the aeronautical radionavigation service, and the radionavigation-satellite service \citep{RR2016,Indermuehle2017TheBETA,ACMA2021}.

~\ac{RAS} has secondary allocation for spectral line observations from \qtyrange{1718.8}{1722.2}{\mega\hertz} and some minimal protection is afforded from \qtyrange{1710}{1930}{\mega\hertz} via footnote \textit{5.149} in the \acp{RR}, stating ``\textit{[..]administrations are urged to take all practicable steps to protect the radio astronomy service from harmful interference}'' \citep{RR2016}. The radio spectrum is therefore a finite resource supporting a diversity of vital services. Current and future radio telescopes operating in large bandwidths outside of the \ac{RAS} protected bands, where usage purely for astronomy purposes is not possible, need to monitor and coexist with other services. 

\ac{ASKAP} is comprised of $36 \times \qty{12}{\metre}$ antennas with a 188-element checkerboard phased array feed at the focus of each dish, allowing up to 36 independent dual-polarisation beams to be formed on the sky \citep{Hotan2021AustralianDescription}. The maximum separation between antenna pairs (baseline length) is approximately \qty{6}{\kilo\metre}. Because low RFI environments are required for radio astronomy, the \ac{ASKAP} telescope was built about \qty{600}{\kilo\metre} north-east of Perth, at Inyarrimanha Ilgari Bundara, the \acs{CSIRO} \acl{MRO}. State and federal legislation has established a protected area known as the \ac{ARQZWA} established by \ac{ACMA} \citep{Wilson2015MeasuresAustralia, Wilson2016TheMeasurements}, a circular region of \qty{260}{\kilo\meter} radius from the observatory centre. However, the regulation does not protect \ac{ASKAP} from orbiting satellites (of particular concern for the planned deployment of several low earth orbit mega-constellations), air traffic navigation, or mobile communication towers outside the \ac{ARQZWA}. The latter of which cause interference during tropospheric ducting events. Ducting is an anomalous propagation phenomenon that can cause the refractive index of the atmosphere to match that of the curvature of the earth, and thereby enables radio waves to refract and travel much farther than usual \citep{Hall1989Radiowave-propagation,Indermuehle2017TheBETA}. 

An holistic strategy to manage the effects of \ac{RFI}, must include legislation,  spectrum management, cooperation, avoidance, flagging, and active \ac{RFI} mitigation  \citep{Hellbourg2012ObliqueAstronomy, Black2015Multi-tierTelescope, Hellbourg2017SpatialArray} but also continued monitoring and evaluation of \ac{RFI}'s impact on the observatory \citep{Indermuehle2017TheBETA}. To this end, observatories and telescopes have started to perform statistical analysis of flagging data \citep{Offringa2015TheMitigation, zhang2021radio, Sihlangu2021Multi-dimensionalObservatories}. In this paper, we present our analysis of \qty{1}{\mega\hertz} resolution flagged data from the \ac{ASKAP} radio telescope's data processing pipeline as a function of frequency and time.
~Section~\ref{sec:background} describes \ac{ASKAP}'s data products and processing as well as the \ac{RFI} environment at the observatory. Section~\ref{sec:implementation} describes the implementation of the flagged statistics pipeline. We present an analysis of the effect of \ac{RFI} on \ac{ASKAP} science, based on over \nhours of \ac{ASKAP} observations in Section~\ref{sec:results}. Finally, future work and concluding remarks are presented in Sections~\ref{sec:future} and ~\ref{sec:conclusion} respectively.

\section{ASKAP data and inferring RFI through flagged data}
\label{sec:background}
The data collected for this project is from processed and archived \ac{ASKAP} continuum observations from \dates. Archived \ac{ASKAP} visibilities are stored with \qty{1}{\mega\hertz} frequency channels and \qty{10}{\second} integration time resolution. \ac{ASKAP} \acp{SSP} typically observe over three ‘clean’ bands, below \qty{1085}{\mega\hertz}, \qtyrange{1293}{1437}{\mega\hertz} (excluding \qtyrange{1149}{1293}{\mega\hertz} severely impacted by radion avigation-satellites which is observed but discarded before processing) and above \qty{1510}{\mega\hertz}, which we refer to as the `clean' low, mid and high bands respectively. For a full description of \ac{ASKAP}, including the technical specifications and overview of science operations, see \cite{Hotan2021AustralianDescription}. 

\ac{ASKAP} operates primarily as an automatically-scheduled telescope to carry out observations, with the autonomous scheduler \acs{SAURON} (\aclu{SAURON}) communicating to the telescope operating system \citep[TOS,][]{Guzman2010TheArchitecture} via parameter specifications \citep[][Moss et al. in prep]{Hotan2021AustralianDescription}. \ac{SAURON} conducts system and environment checks as part of the scheduling decision-making, but \ac{RFI} is largely not included as a factor. The exception to this is ducting avoidance, where some \acp{SSP} have elected to not to observe if there is active ducting due to the impacts on data quality (mainly loss of sensitivity or bandwidth coverage)\footnote{Data from the \acp{SSP} used in this paper do not use ducting avoidance.}. Observations for these \acp{SSP} are not scheduled during active ducting, and if ducting starts during an observation, \ac{SAURON} will interrupt the observation. Further discussion on the severity of ducting is presented in Section \ref{sec:results}. The other most problematic sources of \ac{RFI} are relatively consistent, meaning that active avoidance has not been necessary. The dramatic rise in solar activity approaching the solar maximum (and observed impacts on \ac{ASKAP} calibration and science data) are likely to result in more active consideration of solar activity as part of future upgrades to \ac{SAURON}. Similarly, it has been discussed whether active satellite avoidance as a constraint may be beneficial to especially the mid band observations, but to date the \acp{SSP} and observatory have instead elected to remove the RFI-corrupted lower \qty{144}{\mega\hertz} bandwidth from \qtyrange{1149}{1293}{\mega\hertz}. Work is currently underway as part of the Collaborative Intelligence Future Science Platform collaboration to implement machine learning anomaly detection on \ac{ASKAP} raw data diagnostics. This work will identify and characterise outliers in the data, including \ac{RFI}, and it is expected that the first version of this will be in production before the end of 2023. Note that while \ac{RFI} due to increased solar activity does negatively affect science observations and operations, it does not trigger the flagger and therefore the results presented here are not accounted for by solar interference.

The ASKAPsoft pipeline \citep{Guzman2016StatusOperations, Cornwell2016ASKAPASKAP-SW-0020} is the collection of software required to process \ac{ASKAP} data end-to-end including calibration, ingest, flagging, averaging, imaging, source-finding and archiving. \ac{ASKAP} data is processed at the Pawsey Supercomputing Centre where each of \ac{ASKAP}'s 36 beams are processed in parallel \citep{Hotan2021AustralianDescription}. There are two processes in the pipeline involved in flagging. First, before the data is averaged to \qty{1}{\MHz}, known \ac{RFI} is excised based on an \ac{RFI} database, and then on-the-fly detection and excision occurs \citep{Cornwell2016ASKAPASKAP-SW-0020}. The \ac{RFI} database used, is an internally developed table of known transmitters affecting the observatory \citep{Indermuehle2017TheBETA}. Averaged coarse-resolution \qty{1}{\mega\hertz} data may be flagged again before being imaged and archived \citep{Hotan2021AustralianDescription}. Flaggers generally use fixed and dynamic thresholds in amplitude (or using circular polarisation) to remove anomalous samples in the calibrator and science data. This approach works because many sources of \ac{RFI} have amplitudes much higher than the noise and astronomical signal \citep{McConnell2020TheResults}. Weak noise-like \ac{RFI}, however, can still be missed in the flagging process. The default flagging algorithm embedded in the ASKAPsoft pipeline is \texttt{CFLAG} \citep{CSIRO2022CflagDocumentation}, but the specification parameters also allow the alternative use of \texttt{AOFLAGGER} \citep{Offringa2010Post-correlationMethods, Offringa2012ADetection}. Resulting data products from ASKAPsoft are stored in the \ac{CASDA}\footnote{\href{https://research.csiro.au/casda/}{https://research.csiro.au/casda/}}\citep{Chapman2017CASDA:Archive, Huynh2020TheArchive}. Flagging data is stored alongside visibility data in CASA-style Measurement Sets \citep{Hotan2021AustralianDescription, CASAMs}. ASKAPsoft also outputs diagnostic and quality information throughout the pipeline, including a summary of flagged data per observation.

Our research builds on the existing flag summary reports generated by ASKAPsoft. We establish a baseline of flagged data against which to compare and predict the expected amount of flagged data. We also contribute to the body of research outlining the \ac{RFI} environment at the observatory \citep{Offringa2015TheMitigation,Tingay2020AObservatory,Sokolowski2016TheObservatory,Indermuehle2017TheBETA} and recreate a statistical approach to monitoring of \ac{RFI} flagging which is becoming more widespread in large-scale telescope operations.  \cite{Offringa2015TheMitigation} describes the low-frequency environment at the observatory using ten nights of \ac{MWA} data. Similar approaches at other observatories, include \cite{zhang2021radio} with the \ac{FAST} in China using approximately \qty{300}{\hour} across 45 days, and in South Africa using the MeerKAT telescope where \cite{Sihlangu2021Multi-dimensionalObservatories} implemented a counter-based solution (which we have also used) using approximately \qty{1500}{\hour} of data. Flagging statistics can be used to monitor long-term changes in the \ac{RFI} environment, monitor spectrum compliance and improve the planning of observations such that the impact of \ac{RFI} is minimised. Finally, this work will inform the ongoing development of active \ac{RFI} mitigation \citep{Black2015Multi-tierTelescope, Hellbourg2016InterferenceInterferometry, Chippendale2017InterferenceTelescope} and subsequently monitor the efficacy of these techniques \citep{Hellbourg2012ObliqueAstronomy} in reducing flagging. 

\begin{figure}[ht!]     
    \centering
     \includegraphics[width=\textwidth]{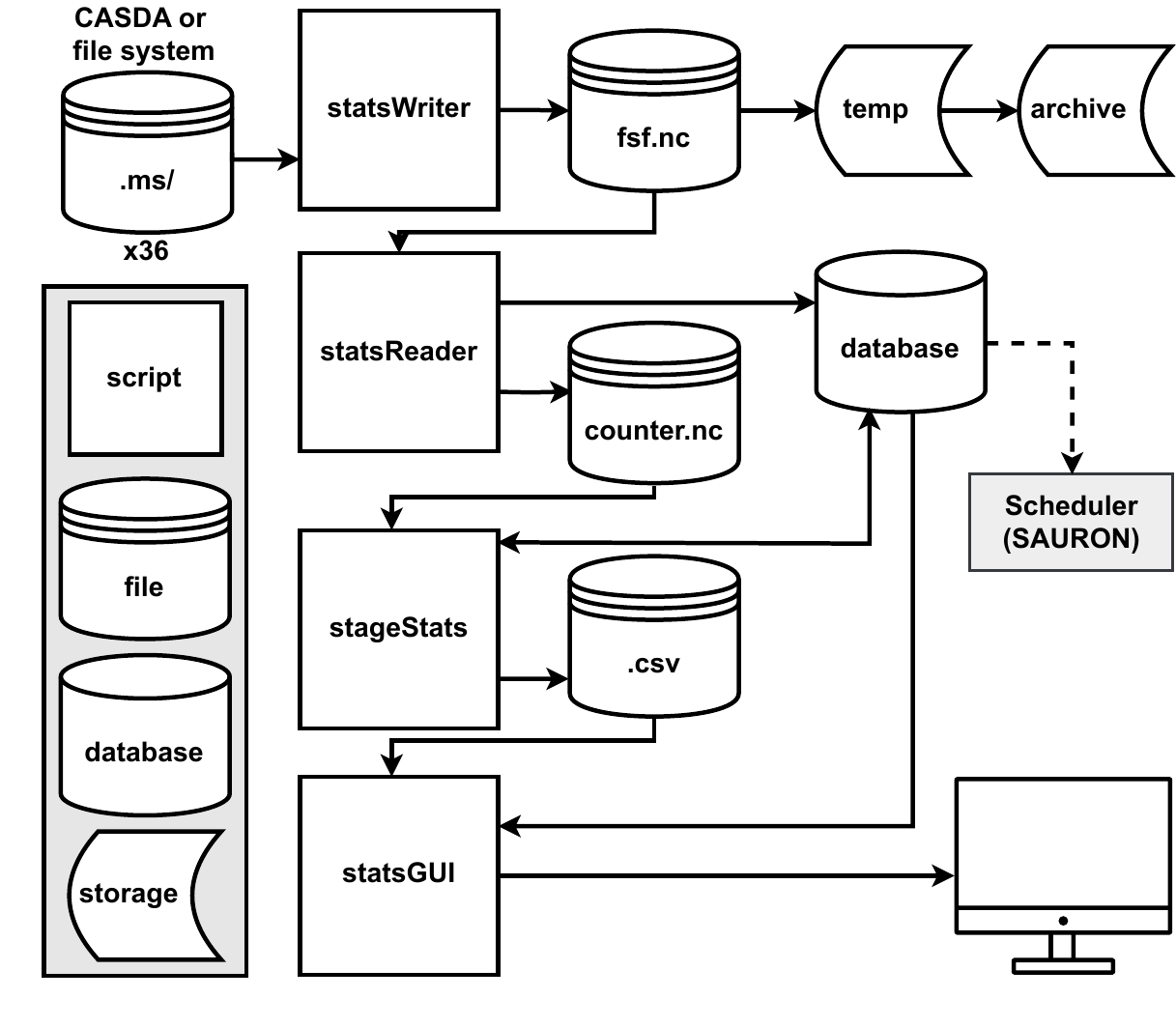}
     \caption{Components of the \texttt{flagstats} library and their data inputs and outputs. \textit{StatsWriter} extracts flags, \textit{StatsReader} aggregates them, \textit{StageStats} further aggregates, performs analysis and prepares files that serve the \acs{GUI}, \textit{statsGUI} for interactive visualisation of flags and aggregated data.}
     \label{fig:sysflow}
\end{figure}

\begin{figure*}[ht!]
\centering
\subfloat[Before cleaning: Horizontal lines show baselines where all frequencies are flagged as a result of a faulty antenna, in this case, antenna 05. Vertical lines show time cycles flagged for reasons other than \ac{RFI}. e.g. slewing.]{\includegraphics[width=0.31\textwidth,valign=t]{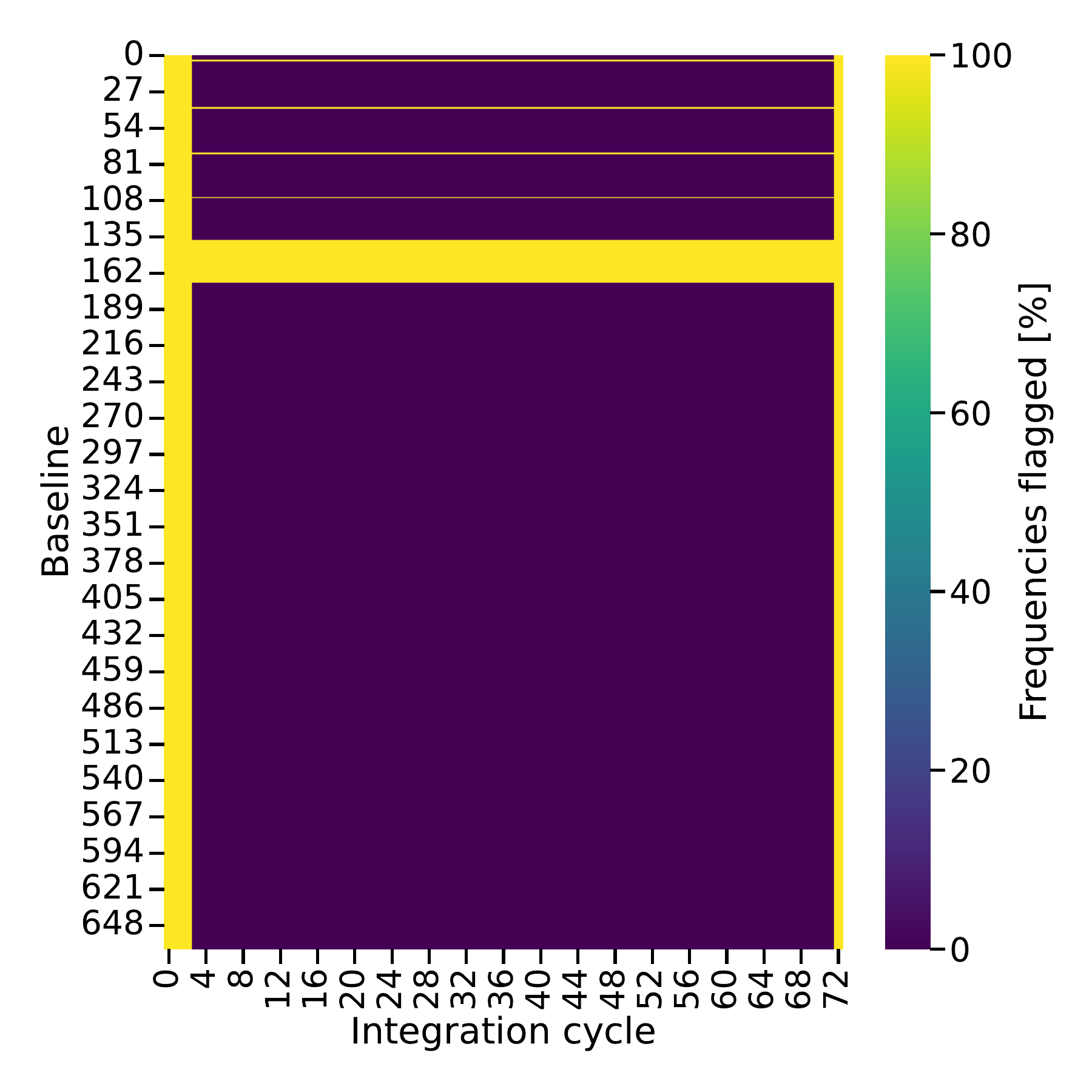}}\hspace{5mm}
\subfloat[Before cleaning: Faulty antennas introduce an offset in the amount of flagged data, note the min value in the colour bar is non-zero. i.e. \qty{5.4}{\percent} of flagging due to other causes.]{\includegraphics[width=0.31\textwidth,valign=t]{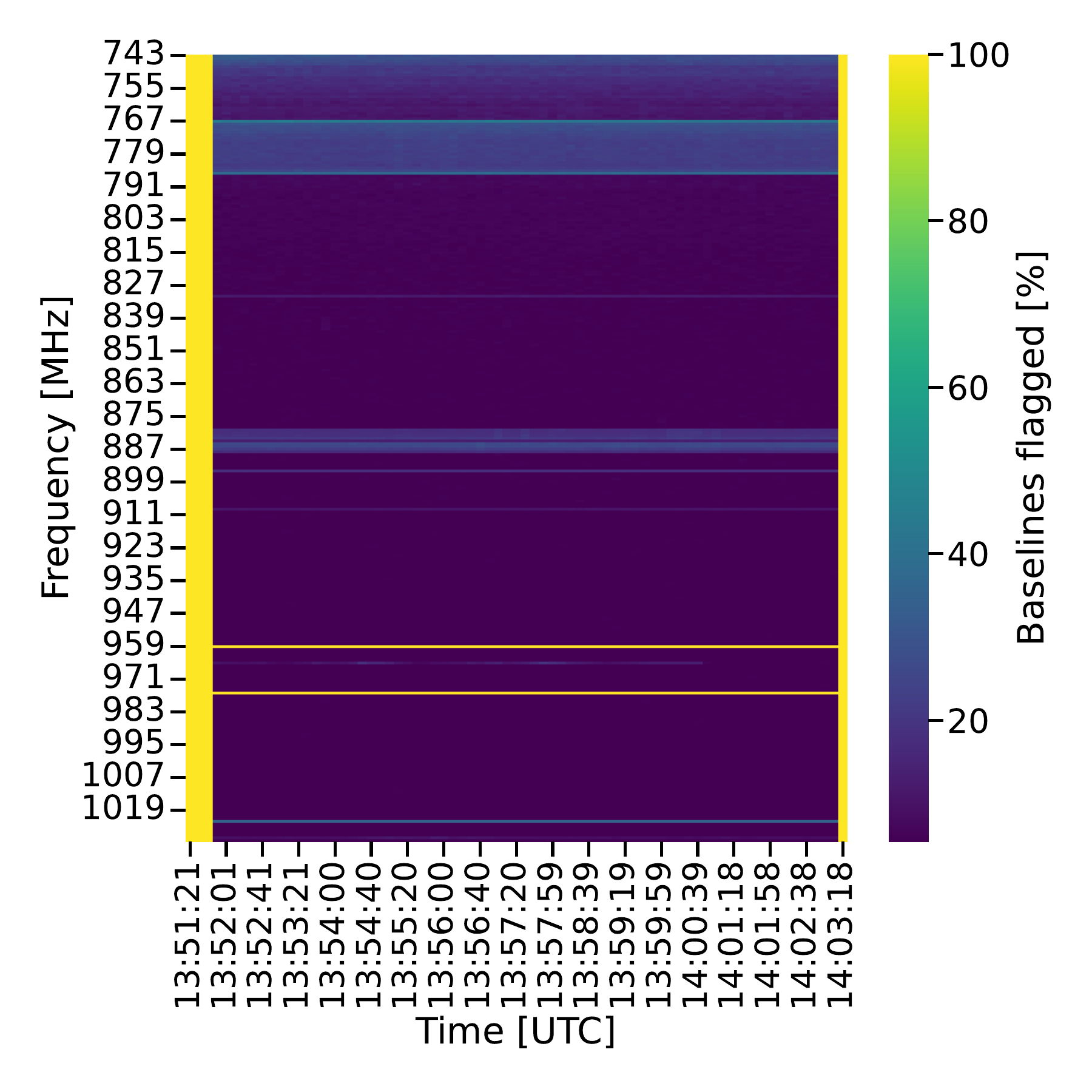}}\hspace{5mm}
\subfloat[After cleaning: Percentage of baselines flagged after removing faulty antennas gives a better estimate of flagged data due to \ac{RFI}. Data will be time-binned into \qty{5}{\minute} bins (overlaid as vertical dashed red lines).]{\includegraphics[width=0.31\textwidth,valign=t]{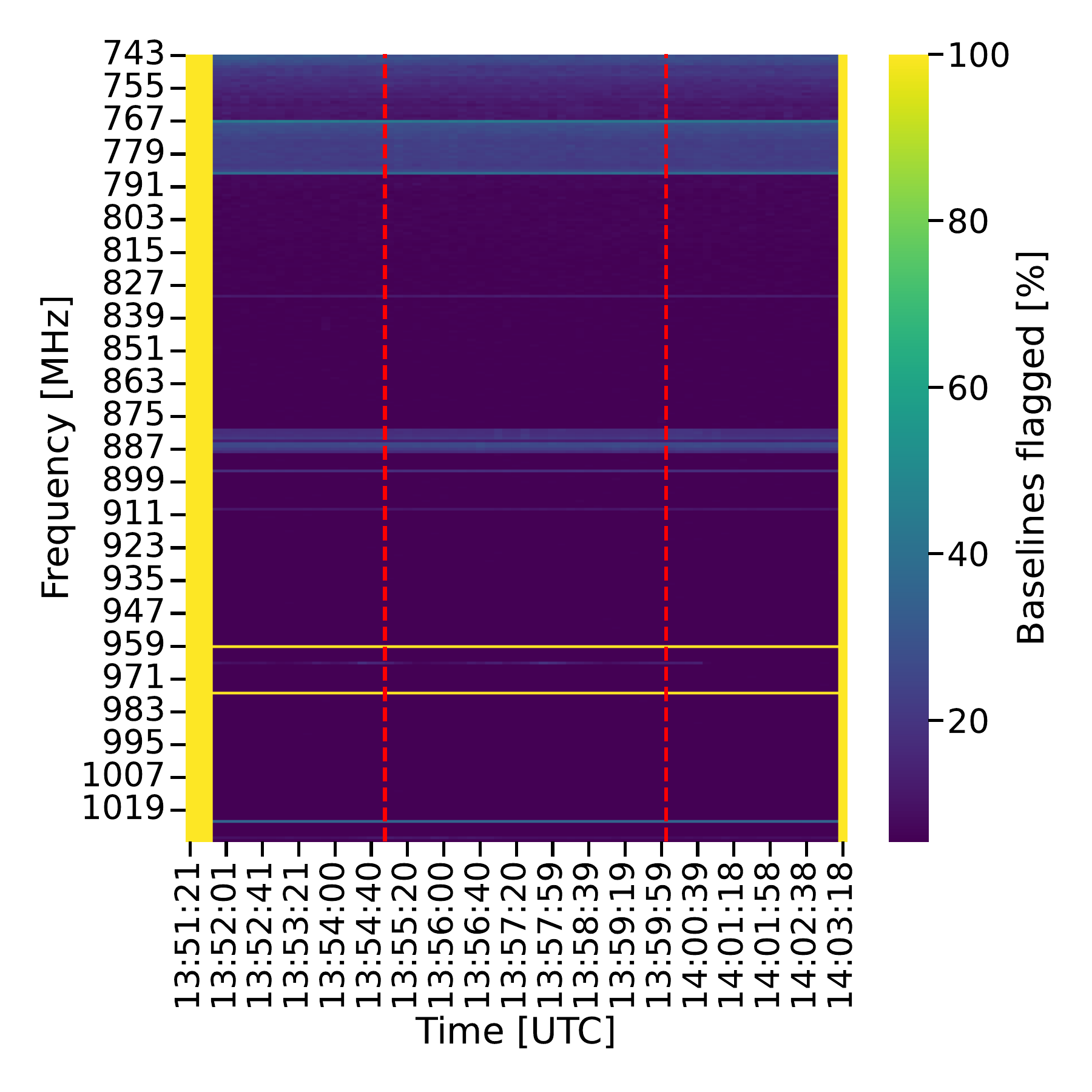}}

\caption{Sub-figures (a) and (b) show the flag cube before removing faulty antennas.  (c) shows the 'cleaned' cube with a more accurate estimate of flags due to \ac{RFI}.}
\label{fig:exampleSBID}
\end{figure*}

\section{Approach and implementation: \texttt{flagstats}}
\label{sec:implementation}
We have used flag tables to infer the effects of \ac{RFI} on \ac{ASKAP} science and to monitor changes in the \ac{RFI} environment as a function of frequency and time.
~We developed software to routinely extract, aggregate, analyse and visualise flagged data from \ac{ASKAP} in near real-time. We opted for a bespoke software solution to systematically aggregate and analyse flagged data because to our knowledge, no compatible open-source software existed. As such, we created the \texttt{flagstats} library based on available Python libraries and existing scripts in \ac{ASKAP}'s pipeline that interrogate \ac{ASKAP} measurement sets and summarise flagged data. The \texttt{flagstats} library is broken into four main components \textit{StatsWriter}, \textit{StatsReader}, \textit{StageStats} and \textit{StatsGUI}. Figure \ref{fig:sysflow} shows the high-level sequence of those components as well as their input and output data products.

\subsection*{StatsWriter}
\textit{StatsWriter} takes a set of \ac{ASKAP} measurement sets as an input and outputs a file containing the flagging tables from each beam
~for the duration of the observation and relevant metadata extracted using \texttt{dask-ms} \footnote{\href{https://dask-ms.readthedocs.io/en/latest/}{https://dask-ms.readthedocs.io/en/latest/}}. Flagging for all four polarisation products (XX, XY, YX, and YY) are recorded in the original measurement set, but, flagging across polarisations is currently the same for \ac{ASKAP}, therefore to reduce the size of the output only the XX polarisation flagging table is stored. A benefit of having these relatively smaller individual \texttt{flagsummary.nc}\footnote{\href{https://www.unidata.ucar.edu/software/netcdf/}{https://www.unidata.ucar.edu/software/netcdf/}} files (per observation) is that users can query them to get comprehensive statistics about the flagging (and \ac{RFI}) affecting a specified observation using \textit{StatsReader}. 

\subsection*{StatsReader}
\textit{StatsReader} is the key component of the \texttt{flagstats} library. It first cleans (removes non-\ac{RFI}-related) flagged data stored in the previous step. A series of `counter' files are created if they do not exist. Data is binned for each observation, in turn. The appropriate counter is updated to aggregate flagged data from multiple observations. Counters from which statistics are computed are updated with the flagged and total number of contributing samples of each bin similar to \cite{Sihlangu2021Multi-dimensionalObservatories}. 

Consider a single beam from an \ac{ASKAP} observation, where the flag table from the measurement set is reshaped and stored into a 3-dimensional array of shape, $n_\text{freq} \times n_\text{cycles} \times n_\text{baselines}$. Figure \ref{fig:exampleSBID}a shows the percentage of flagged frequency channels in the observation as a function of time (in integration cycles) on the x-axis and baseline on the y-axis. Figure \ref{fig:exampleSBID}b, is the same cube but shows the percentage of flagged baselines in the observation as a function of time (in \acs{UTC}) on the x-axis and \qty{1}{\mega\hertz} frequency channels on the y-axis. We call Figure \ref{fig:exampleSBID}a and Figure \ref{fig:exampleSBID}b the `dirty' cube because faulty and intermittent antennas have not yet been removed. \ac{ASKAP} currently uses boolean flags which are \texttt{True} when the data is discarded/flagged. One of \textit{StatsReader}'s functions is to identify (and remove) flags that are not due to \ac{RFI}. The vertical yellow lines in Figure \ref{fig:exampleSBID}a show time cycles for which all frequency channels are flagged, similarly horizontal lines show baselines for which all frequency channels are flagged throughout the observation. For the latter, an antenna can be mapped to a set of baselines due to a faulty antenna(s), in this case antenna ak05. For the former, flags are from timing issues (slewing time if all antennas are not in the `tracking' state and latency of data flow through the digital systems). \textit{StatsReader} will disregard flags where all frequencies are affected over either all baselines or all integration cycles because they are not from RFI. \textit{StatsReader} also cleans antennas that `drop out' mid-observation.

Figure \ref{fig:exampleSBID}c shows the same cube but masked, i.e. without the bad integration cycles and antennas, that we refer to as the `clean' cube, which yields a better upper estimate of the flags due to \ac{RFI}. Using the Radio Regulations \citep{RR2016}, spectrum plan \citep{ACMA2021}, and known \ac{RFI} affecting \ac{ASKAP} \citep{Indermuehle2017TheBETA} we can map features in Figure \ref{fig:exampleSBID}c to likely sources of \ac{RFI}. It should be noted however that the \qty{1}{\MHz} frequency resolution is coarse compared to the bandwidth of many interferers, limiting the accuracy with which we can map flags to services. Furthermore, multiple interferers may also affect the same \qty{1}{\MHz} channel. Brighter horizontal lines in Figure \ref{fig:exampleSBID}c show channels for which more (or all remaining) baselines/antennas are affected by \ac{RFI}. Two red vertical dashed lines demarcate the three bins that this observation would be divided into. Starting from the top of the plot, this \qty{12}{\minute} observation is corrupted by mixed fixed cellular interference below \qty{787}{\MHz} \citep{Indermuehle2017TheBETA}. In the frequency range of this observation, there are many mobile base stations that routinely trigger \texttt{CFLAG} due to ducting, discussed further in Section \ref{sec:results} (see Figure \ref{fig:basestationMap_Balt}). For example, we know of two Vodafone base stations (identified by decoding \ac{RFI} monitoring equipment) that operate in a frequency range of the affected channels in Figure \ref{fig:exampleSBID}c \citep{Indermuehle2017TheBETA}: the first in the Northampton Shire, \qty{267}{\kilo\metre} south-west of \ac{ASKAP} and the second in Carnarvon, \qty{357}{\kilo\metre} north-west of \ac{ASKAP}. In this case \qty{960}{\MHz}, like the \qty{976}{\MHz} flagged channel above it, the origin of the \ac{RFI} is likely self-generated. 

\begin{figure*}[!hbt]
    \centering
    \includegraphics[width=0.95\textwidth]{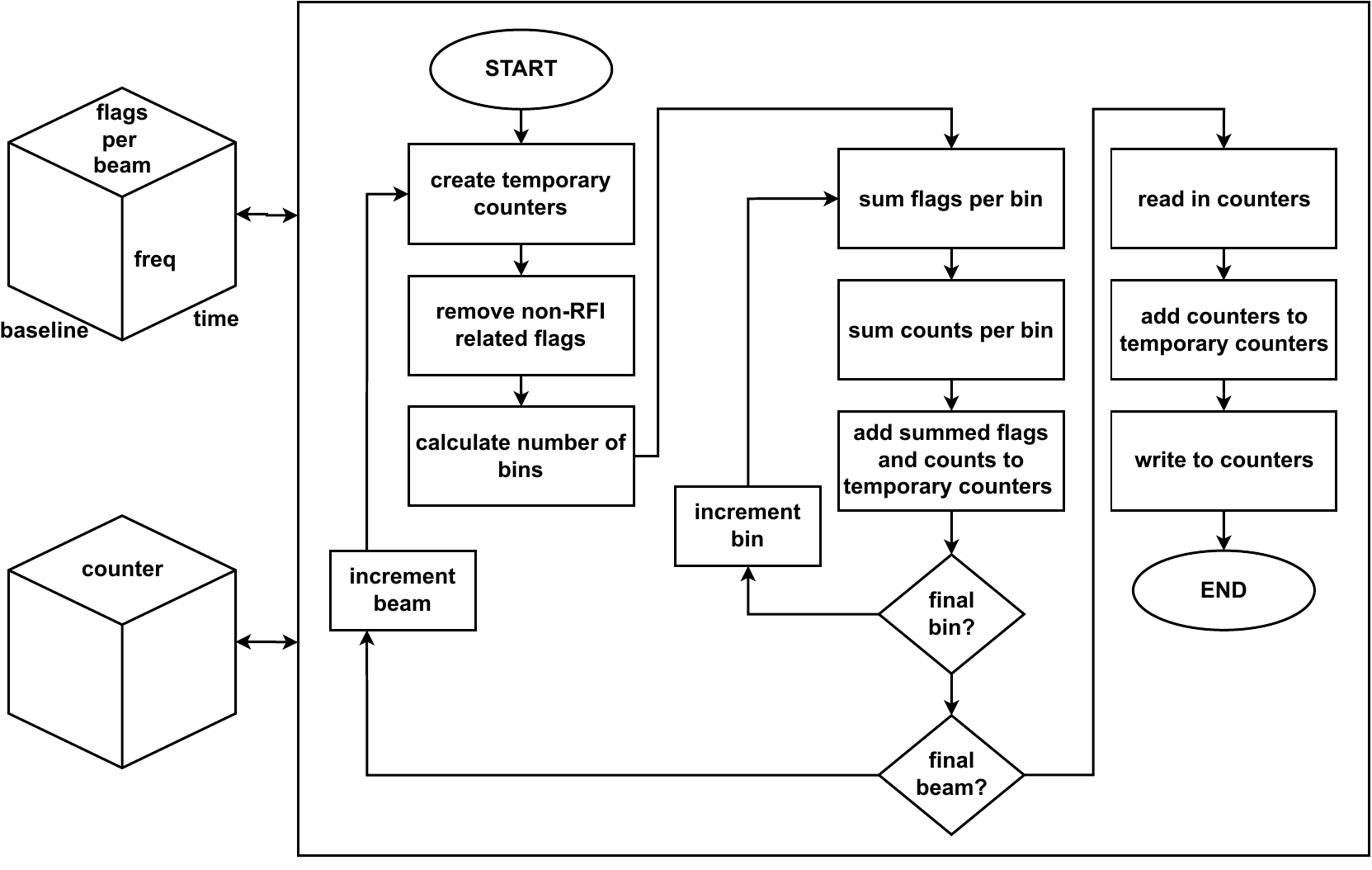}
    \caption{\textit{StatsReader} takes an array of each beam's flags in turn, creates a counter if one does not already exist, cleans the flag cube, bins the data, and updates the appropriate counter. This process is repeated for every observation. There are counters for all permutations based on the metric calculated, year and data source (science, bandpass, \ac{RFI} survey). Science data is further divided by the different \acfp{SSP} processed (\acs{EMU}, \acs{POSSUM}, \acs{RACS}, \acs{VAST}).}
    \label{fig:readerflow}
\end{figure*}

An overview of the steps involved in \textit{StatsReader} for an arbitrary counter are depicted in Figure \ref{fig:readerflow}. Each \texttt{counter.nc} file has three groups\footnote{\href{https://docs.xarray.dev/en/stable/}{https://docs.xarray.dev/en/stable/}}, one for each of the following counters:
\begin{itemize}
    \item The \texttt{flag} counter, tallies flags flagged as \texttt{True}, per bin.
    \item the \texttt{count} counter, tallies the total number of samples per bin, i.e. flagged as \texttt{True} or \texttt{False}.
    \item The \texttt{int\_time} counter, scales the count by the integration time of each observation to keep track of the flagging per unit time (typically represented in hours but stored in minutes).
\end{itemize}

We have implemented a series of counters as opposed to a single counter \citep{Sihlangu2021Multi-dimensionalObservatories}. First, for each metric we want to interrogate, there are individual counter files:

\begin{itemize}
    \item Time of Day, arrays in this counter have shape $1100 \times 666 \times 288$: $1100 \times \qty{1}{\mega\hertz}$ frequency bins, 666 baseline pairs from \ac{ASKAP}'s 36 antennas and $288 \times \qty{5}{\min}$ time bins in a day. Figure \ref{fig:exampleSBID}c shows these bins with two red vertical dashed lines.
    \item Day of Week, arrays in this counter have shape $1100 \times 666 \times 168$; frequency, baseline and 168 $\times$ \qty{1}{\hour} time bins in a week.
    \item Week of Year, arrays in this counter have shape $1100 \times 666 \times 53$; frequency, baseline, 52 complete weeks in a year and one incomplete week. Week zero is the week inclusive of the 4\textsuperscript{th} of January using the ISO week date system.
\end{itemize}

The above set of counter files is created for each permutation of year and type of data. \acp{SSP} are further divided by the different surveys processed: \acs{EMU}, \acs{POSSUM}, \acs{RACS}, \acs{VAST} (see Section \ref{sec:results}). \textit{StatsReader} uses metadata stored in the extracted flags by \textit{StatsWriter} to locate and update the appropriate counter files. This approach makes it easy to add new metrics, types of data or surveys by simply adding the associated counters. In addition to aggregating, cleaning, binning and updating counters, Figure \ref{fig:sysflow} shows that \textit{StatsReader} writes high-level data, per observation to a database table. This table is used to keep track of the observations processed and includes the percentage of total flagged data (including bad antennas), percentage of flagged data due to \ac{RFI}, duration, field of observation, data source and survey name. At this point, counters per metric for each observation are aggregated in their respective counters - by year, type of data and science survey - but not aggregated with each other.

\subsection*{StageStats}
In order to create aggregated statistics for each metric, across different years, surveys and observing modes, counters of the same type must be added together. First, vertically, the counters created and updated by \textit{StatsReader} in the lowest directory of the file hierarchy are added together by \textit{StageStats} and placed at the level of the file structure above it until there is one counter per metric per year. Then counters are added horizontally until there is one counter per metric per type of data. Finally, counters from every year are added together. \textit{StageStats} has three other processes: 
\begin{itemize}
    \item Create tables from these `top-level' counters and store them in the database with local copies to serve the \ac{GUI}. 
    \item Calculate high-level statistics based on the files processed by \textit{StageReader} (Table \ref{tab:survey_distribution}). We also use this database table to infer observations affected by ducting, based on excess flagging. 
    \item Perform analysis of the aggregated data (see Section \ref{sec:results}):
    \begin{itemize}
        \item Compare flagging between day and night (Figure \ref{fig:day_vs_night}).
        \item Aggregate data from \ac{RFI} monitoring equipment.
        \item Compare flagged and \ac{RFI} monitoring equipment data.
        \item Map flagged data to corresponding radio service(s).
        \item Calculate losses per service (Figure \ref{fig:EBL}).
        \item Calculate losses per service year-on-year (Figure \ref{fig:year_on_year}).
        \item Infer observations affected by ducting based on `excess' flagging.
    \end{itemize}
\end{itemize}

Finally, \textit{StageStats} summarises statistics per metric (time of day, week and year 
), to reduce the computation required for the interactive GUI. 

\subsection*{StatsGUI}
The \textit{StatsGUI} is an interactive \ac{GUI} that allows users to explore aggregated flagged data. It also matches \ac{RFI} probabilities to radio services and known transmitters in the same \qty{1}{\MHz}~ channel using a click callback. Users first select a metric to view (by default it is `time of day'), they can then use a series of filters to select data from a particular year, survey and/or observation type. We are in the process of automating the individual parts of the \texttt{flagstats} library. So far \textit{StatsWriter} is fully implemented to extract flags from all bandpass and continuum science observations as soon as the processing is completed. 

\subsection*{Automation}
The \ac{ASKAP} telescope is increasingly becoming an autonomous system building upon \ac{SAURON}, a dynamic observation scheduling system (Moss et al. in prep), and \texttt{Process Manager}, an event-based triggered workflow manager. \texttt{Process Manager} has been designed to allow the automatic triggering of a data processing pipeline (e.g. calibration and imaging) when a set of defined criteria have been met. We used the \texttt{Python} package \texttt{prefect}\footnote{\url{https://github.com/PrefectHQ/prefect}} to create an extensible workflow that implements the \textit{StatsWriter} routines, which was then registered with \texttt{Processing Manager} and configured to trigger against all science-observations once they have been processed by the larger \ac{ASKAP} processing-pipeline. Hence, the flagging statistics that are currently extracted are from the calibrated measurement sets after the automated \ac{RFI} flagging within the \ac{ASKAP} processing pipeline have been performed. We plan to further develop our own \texttt{prefect}-based pipeline to directly run flagging procedures against the raw uncalibrated measurement set before they are processed by software in the larger \ac{ASKAP} ecosystem. This improvement will gather useful statistics on sub-bands that are currently removed at the beginning of the \ac{ASKAP} pipeline containing excessive \ac{RFI} and therefore affecting scientific processing, for example over the frequency range \qtyrange{1149}{1293}{\mega\hertz} impacted by the radionavigation-satellite service. While we continue to test, develop, optimise and automate the software it remains internal, but we intend to explore options to generalise and make it more widely available as we further develop the software. In the next section, we present the preliminary results based on \qty{1}{\mega\hertz} flagged data.

\begin{figure*}[!ht]
  \includegraphics[width=\textwidth]{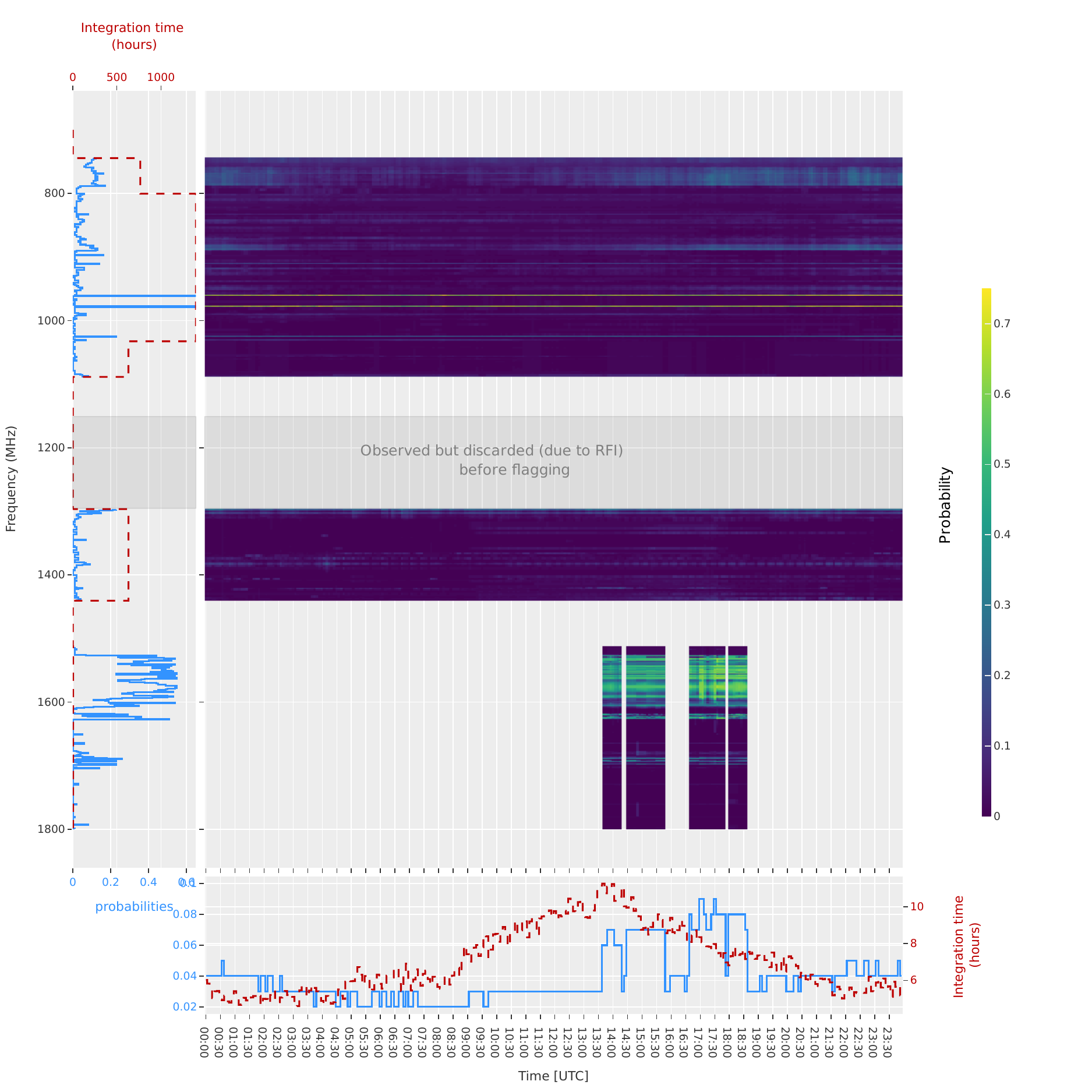}
  \caption{The central heatmap shows the probability of flagging due to \ac{RFI} based on the aggregation of over \nSBIDs scheduling blocks comprising approximately \nhours of observing time from \dates. The x-axis is binned to \qty{5}{\minute}s, and the y-axis to \qty{1}{\MHz} frequency channels. The vertical subplot to the right shows the average flagging due to \ac{RFI} in blue as a function of frequency. Similarly, the horizontal subplot at the bottom shows the average flagging due to \ac{RFI}  as a function of the time of day in blue. The red curves in both subplots show the number of hours that contributed data to that bin. Three sub-bands in which \ac{ASKAP} continuum observations are most typically conducted are shown. From top to bottom, the low band shows flagging due to \ac{RFI} from the fixed/mobile service exacerbated by ducting, the two most prominent peaks are likely due to self-generated interference in \ac{ASKAP}'s signal chain. Aeronautical mobile and radionavigation affect the low and mid bands shown above. Note the bottom half of the mid band from \qtyrange{1149}{1293}{\mega\hertz} whilst observed is not processed or archived as it is severely affected by radionavigation-satellites. The lower half of the high band is affected by various satellite services, in addition the high band is affected by meteorlogical aids and satellite services as well as fixed/mobile. Both the mid and high bands have protected \acl{RAS} allocations for HI and OH spectral lines respectively.}
  \label{fig:HOD}
\end{figure*}

\section{Results}
\label{sec:results}
In this paper, we present results based on science observations with at least \qty{250}{\hour} per year from \dates\footnote{2021 and 2023 have at least \qty{750}{\hour} and \qty{500}{\hour} respectively.}. 
Flags from archived science data were collected via \ac{CASDA}. Recent science data were directly processed from the Pawsey Supercomputing Centre file system. Most recently this is done automatically by triggering \textit{StatsWriter} upon completion of the \ac{ASKAP} processing pipeline. Table \ref{tab:survey_distribution} shows the total observing time  and number of observations of science data subdivided by survey and clean band.  Science data from the following \acp{SSP} and their respective pilot surveys have been included: 
\begin{itemize}
    \item the \ac{RACS} I, II and RACS-mid \citep{McConnell2020TheResults, Hale2021TheRelease, Duchesne2023TheRACS-mid}, 
    \item the \ac{VAST} \citep{Murphy2013VAST:Transients, Murphy2021TheSurvey}, 
    \item the \ac{POSSUM} \citep{Anderson2021EarlyCluster} and, 
    \item the \ac{EMU} \citep{Norris2011EMU:Universe}     
\end{itemize} 

\begin{table}[!ht]
\caption{Hours of continuum science data by processed \acl{SSP} and band.}
\begin{threeparttable}
\begin{tabular}{@{}lrrrr@{}}
\toprule
Project &
  Typical Field &
  Clean &
  \# Fields &
  Total \\
   &
  Duration &
  Band &
   &
  Hours \\\midrule
    RACS & 
    \qty{0.25}{\hour} & 
    \begin{tabular}[c]{@{}r@{}} low \\ mid \\ high\end{tabular} & 
    \begin{tabular}[c]{@{}r@{}} 864 \\ 1574 \\ 18\end{tabular} & 
    \begin{tabular}[c]{@{}r@{}} 385 \\ 390 \\ 4\end{tabular}\\ \hline
    VAST & 
    \qty{0.20}{\hour} & 
    \begin{tabular}[c]{@{}r@{}} low \\ mid\end{tabular} & 
    \begin{tabular}[c]{@{}r@{}} 2146 \\ 429\end{tabular}& 
    \begin{tabular}[c]{@{}r@{}} 50 \\ 36\end{tabular}\\ \hline
    POSSUM & 
    \qty{10}{\hour} & 
    \begin{tabular}[c]{@{}r@{}}low \\ mid\end{tabular} & 
    \begin{tabular}[c]{@{}r@{}}1 \\ 10\end{tabular} & 
    \begin{tabular}[c]{@{}r@{}}10 \\ 98\end{tabular}\\ 
    EMU & 
    \qty{5}{\hour} or \qty{10}{\hour} & 
    low & 
    77 & 
    617\\  \bottomrule
\end{tabular}
\end{threeparttable}
\label{tab:survey_distribution}
\end{table}

A key outcome of this work is shown in Figure \ref{fig:HOD}. Using the flagstats library, the data in Table \ref{tab:survey_distribution} was extracted, cleaned, binned, aggregated, summed across \ac{SSP}, data category and metric and analysed. Figure \ref{fig:HOD} shows the probabilities of flagging due to \ac{RFI} as a function of frequency and time of day. The x-axis, binned to \qty{5}{\minute}, is given in \ac{UTC} in the format \texttt{hh:mm}. 

The y-axis shows \ac{ASKAP}'s \qty{1}{\mega\hertz} resolution frequency channels in ascending order. The vertical subplot to the left shows the average flagging due to \ac{RFI} across the day as a function of frequency in blue. Similarly, the horizontal subplot at the bottom shows the average flagging due to \ac{RFI} across the ASKAP band as a function of the time of day in blue. Both subplots show the contributing number of observing hours per bin in red.

An upper estimate on the total percentage of flagged data due to \ac{RFI} across \ac{ASKAP}'s frequency range is \percentFlagged. Starting from the lowest frequencies and working upwards, we will discuss  the corresponding primary service allocations \citep{ACMA2021} to features in Figures \ref{fig:HOD} and \ref{fig:day_vs_night}. %

Figures \ref{fig:HOD} and \ref{fig:day_vs_night} show flagging below \qty{787}{\MHz}, and around \qty{841}{\MHz}, \qty{884}{\MHz} and \qty{894}{\MHz} which are most likely attributable to the closest mobile base stations operating at those frequencies (see Figure \ref{fig:basestationMap_Balt}). Flagged data above \qtyrange{960}{1163}{\mega\hertz} range is likely due to aeronautical radionavigation. The end of \ac{ASKAP}'s clean low band (\qty{1085}{\MHz}) terminates where \ac{RFI} due to aeronautical radionavigation is expected to increase based on measurements by \ac{RFI} monitoring equipment (middle panel, Figure \ref{fig:day_vs_night}). 

Radionavigation-satellite routinely affect the \qty{144}{\MHz} below \qty{1293}{\mega\hertz}, although observed, this sub band is discarded before processing result in a large section of bandwidth in which no flagging data has been collected. The clean mid band from \qtyrange{1293}{1437}{\mega\hertz} is affected by flagging in corresponding to the following services: radio location service for use by the Australian Defence Force and Department of Defence from \qtyrange{1300}{1400}{\mega\hertz} and the \acl{RAS} hydrogen line protected band from \qtyrange{1400}{1427}{\mega\hertz} in which the strong hydrogen line emission itself may be triggering \texttt{CFLAG}. 

Various satellite services affect the bottom half of the high band (\qtyrange{1510}{1797}{\mega\hertz}) including radionavigation-satellite satellite and radiodetermination-satellite services as well as aeronautical radionavigation. The \acl{RAS} OH spectral line protected band from \qtyrange{1610.6}{1613.8}{\mega\hertz} and \qtyrange{1660}{1670}{\mega\hertz} is located between the above satellites and meteorlogical aids and satellite services and fixed/mobile which fills the remainder of \ac{ASKAP}'s frequency range.

\subsection*{ASKAP self-generated RFI}
\label{sec:self-generated}

Not all \ac{RFI} is external to the observatory and telescope, with \ac{ASKAP} itself (and the infrastructure supporting it) capable of generating \ac{RFI}, which might include power supplies and electronic components. Self-generated \ac{RFI} is generally unintended and we aim to identify and ultimately reduce the impact of these unwanted sources of \ac{RFI} caused within the observatory. 

An example of this in early \ac{ASKAP} observations/operations is the \ac{ODC} a noise source used for calibrating the \ac{PAF} which introduced \ac{RFI} (particularly in the shortest, baselines and in the low part of the band). The \ac{ODC} is now only switched on as needed, during the beginning of beamforming and before an observation to update the weights, not during the observation. The \ac{ODC} is designed to be broad-band and operates at about \qty{1}{\percent} of the system temperature. It is therefore unlikely to have caused any additional flagging. Furthermore, the majority of data presented herein was collected after this issue was resolved. Shielding is also used to mitigate the effects of \ac{RFI} in the observatory control building. In some cases due to planned activities at the observatory (including maintenance and construction), the risk of \ac{RFI} is increased. In these instances a \ac{REMP} is conducted to prevent and avoid \ac{RFI} as much as possible. 

Another example includes some digital artifacts that impact the data like \ac{RFI} but can be coherent across the array without being radiated into it due to their originating from coherent clock signals. A dominant example is a \qty{256}{\mega\hertz} clock that is multiplied by 32/27 to read out the digital receiver’s \qty{1}{\MHz} resolution oversampled coarse filter bank \citep{Tuthill2012DevelopmentASKAP, Brown2014DesignSystem}. Table \ref{tab:birides} shows calculated values of harmonics of this clock signal and frequencies that they will appear at in observations due to the direct sampling architecture of \ac{ASKAP}'s digital receiver.  This clock signal is narrowband and correlated across antennas.

\begin{table}[h]
\caption{Calculated interference due to harmonics from $32/27 \times \qty{256}{\MHz}$ coarse filterbank readout clock in \ac{ASKAP}'s digital receiver. Bolded frequencies are potentially triggering the flagger, strikethrough text indicates frequencies outside the available \ac{ASKAP} observing band. Apparent frequencies calculated for direct sampled filter bands as defined in \citep{Brown2014DesignSystem}.}
\label{tab:birides}
\begin{tabular}{rrrr}
\hline
\multicolumn{1}{r}{\textbf{Harmonic}} & \multicolumn{3}{c}{\textbf{Frequency (MHz)}}                                                              \\ 
\multicolumn{1}{c}{\textbf{}}         & \multicolumn{1}{r}{\textit{\textbf{RF}}} & \multicolumn{1}{r}{\textit{\textbf{1200 MHz Filter}}} & \multicolumn{1}{r}{\textit{\textbf{1450 MHz Filter}}} \\ \hline
1 & 303.41& \textbf{976.59}  & 1232.59 \\
2 & 606.81& \st{673.19}  & 929.19  \\
3 & 910.22& \textbf{910.22}  & \textbf{910.22}  \\
4 & 1213.63& 1213.63 & 1213.63 \\
5 & 1517.04& 1042.96 & 1517.04 \\
6 & 1820.44& 739.56  & 1251.56 \\
7 & 2123.85& 843.85  & \textbf{948.15}  \\
8 & 2427.26& 1147.26 & 891.26  \\
9 & 2730.67& 1109.33 & 1194.67 \\ \hline
\end{tabular}
\end{table}

This signal is internal to each digital receiver card and is well shielded inside \ac{ASKAP}'s shielded room. It enters the signal path directly on the digitiser cards. Note the \qty{673.19}{\mega\hertz} value has been struck through because it is outside of \ac{ASKAP}'s observing range and therefore does not see corresponding flagging in the data collected herein.

\begin{figure}[!ht]
  \includegraphics[width=\textwidth]{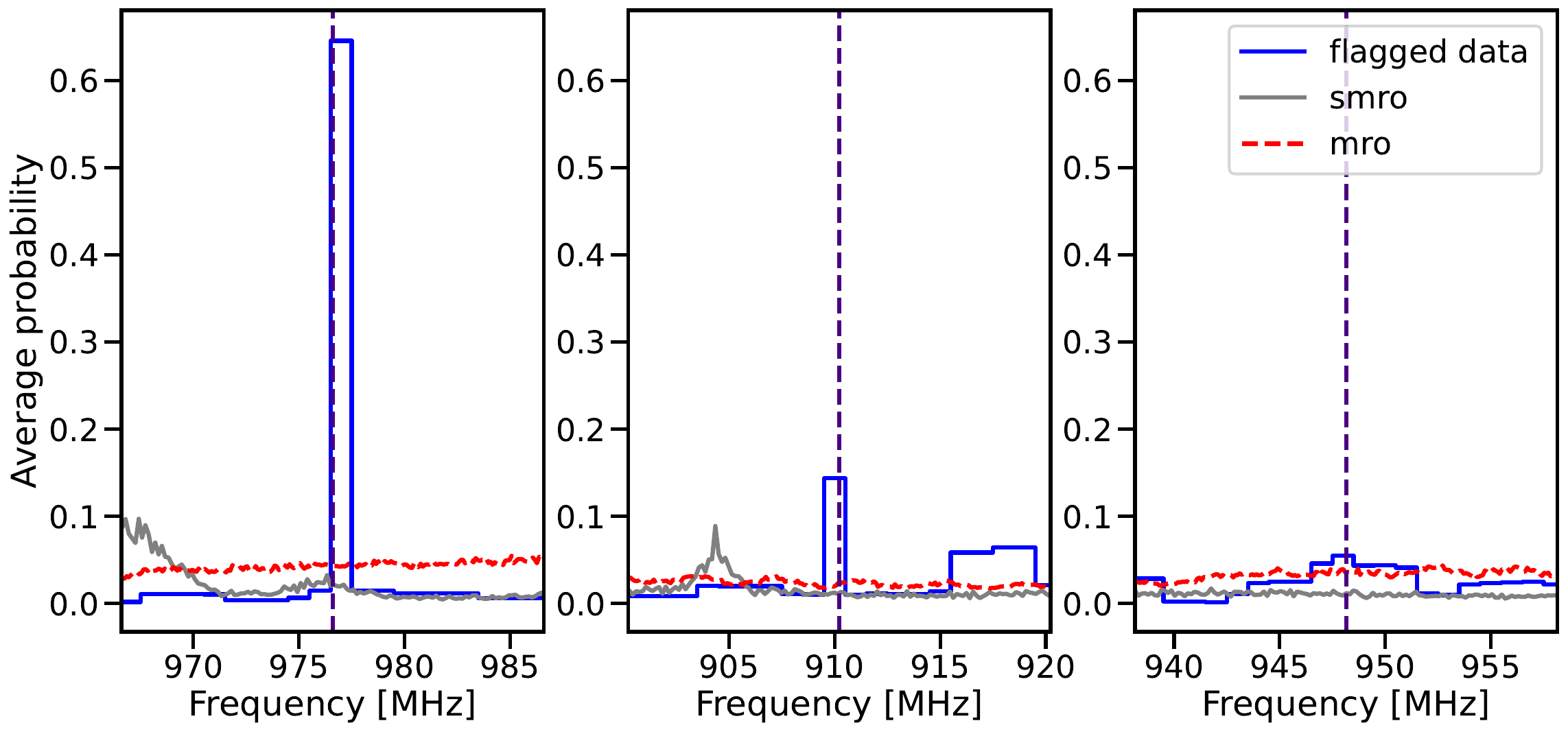}
  \caption{Mapping flagged data at the calculated harmonics of self-generated interference, overlaid are the probabilities of \ac{RFI} based on the \ac{RFI} monitors. For these frequencies there is no externally measured \ac{RFI} but flagging, consistent with self-generated interference.}
  \label{fig:self-generated}
\end{figure}

Figure \ref{fig:self-generated} shows how by using the values in Table \ref{tab:birides} we map the probabilities of flagging due to \ac{RFI} to frequencies that are likely flagged due to self-generated interference. i.e. the bolded frequencies in Table \ref{tab:birides} are potentially triggering the flagger. The flagged data from Figure \ref{fig:HOD} is shown in Figure \ref{fig:self-generated} as a solid blue step plot, with step widths of \qty{1}{\MHz}. Overlayed and with a higher frequency resolution, are two separate \ac{RFI} monitors used at the observatory, the first (solid grey) and the second (dashed red). The first is more sensitive owing to its slower scanning rate and thus longer integration time (`s' stands for sensitive in the name). It is better suited for detection of faint signals often seen from satellites, distant transmitters such as mobile base stations (See Figure \ref{fig:day_vs_night}), and localised weak electromagnetic interference. The faster and thus less sensitive \ac{RFI} monitor is paired with an active antenna to make up in part for the faster scan rate. Both scan from \qtyrange{20}{3000}{\mega\hertz}, the first has finer frequency resolution, \qty{3.125}{\kilo\hertz}, with a sweep time of \qty{62}{\second}, compared to the second which scans at \qty{100}{\kilo\hertz} with a sweep time of \qty{1.49}{\second}, permitting detection of rapid transient emissions. More information on \ac{RFI} monitoring equipment can be found in \cite{Indermuehle2017TheBETA}. The point is that at the frequencies with flagging peaks in Figure \ref{fig:self-generated} and no corresponding measurement by the \ac{RFI} monitoring equipment, we detected three self-generated \ac{RFI} candidate frequencies on which to follow up. There is a mechanism in the \ac{ASKAP}soft \texttt{injest} software, `on-the-fly averaging' based on an amplitude threshold, that should remove the underlying narrowband signal causing \texttt{CFLAG} to flag these \qty{1}{\MHz} channels; we believe these candidates are not strong enough to trigger this mechanism and are therefore subsequently flagged. An additional candidate signal of this nature not previously documented, the \qty{960}{\MHz} flagged channel in Figure \ref{fig:HOD} and \ref{fig:day_vs_night} has been identified for further investigation.

\subsection*{RFI-affected science data}
The elevated steps in the red vertical subplot of total integration time in Figure \ref{fig:HOD} (where we have accumulated more data) identify two clean bands that \ac{ASKAP} regularly observes in the low and mid bands. Across the science bands flagging due to \ac{RFI} is \qty{3}{\percent}. 

The average flagging in the low band is \qty{5}{\percent} and is mostly accounted for by four  primary service allocations. 

\begin{figure}[!ht]
  \includegraphics[width=\textwidth]{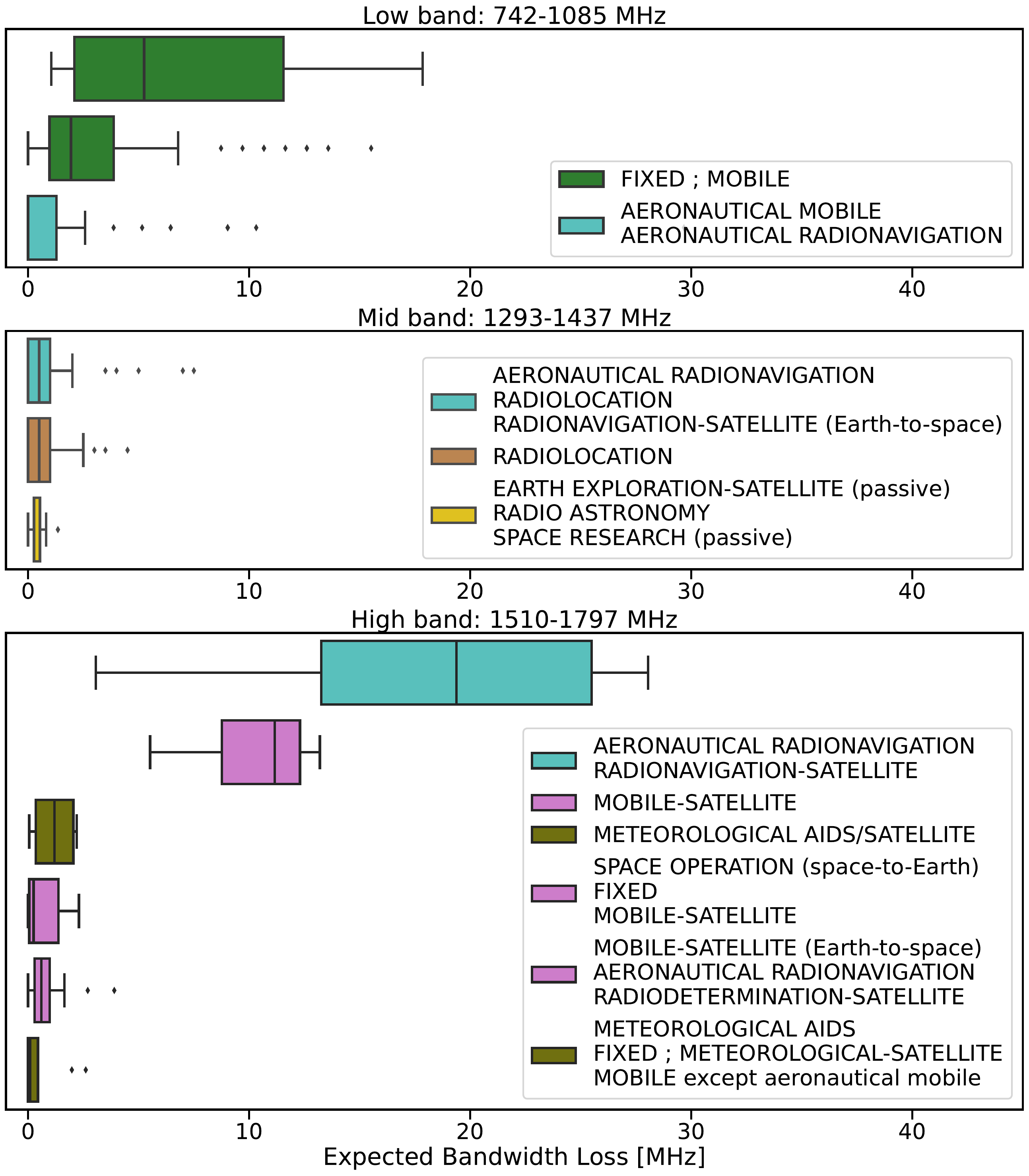}
  \caption{An estimated bandwidth loss is calculated by multiplying the channel width (\qty{1}{\mega\hertz}) by the probability of flagging in that channel and summing all the channels corresponding to a particular primary service allocation. For each band --- \ac{ASKAP}'s low (top panel), mid (middle panel) and high (bottom panel) bands --- the estimated bandwidth loss of the highest average flagged services are shown. The mid band (excluding radionavigation-satellite affected data) is the least affected by \ac{RFI} followed by the low and high bands. The low band is most affected in fixed/mobile (green) allocations and the high band by aeronautical (cyan) and satellite allocations (pink).}
  \label{fig:EBL}
\end{figure}

Less than \qty{0.1}{\mega\hertz} is lost to \ac{RFI} in this band without identification of a corresponding primary service allocation. The top panel of Figure \ref{fig:EBL} shows the two primary service allocations that most affect the low band are both fixed/mobile allocations (green) to which combined \ac{ASKAP} science loses between \qty{6}{\mega\hertz} and \qty{20}{\mega\hertz}. Estimated losses are calculated by multiplying the channel width (\qty{1}{\mega\hertz}) by the probability of flagging in that channel and summing all the channels corresponding to that service resulting in a range of expected/estimated bandwidth losses in MHz. 

The average mid band flagging is \qty{1}{\percent}. This does not include the \qty{144}{\mega\hertz} below \qty{1293}{\mega\hertz} discarded before processing \qty{100}{\percent} of the time which is affected by radionavigation-satellites. Including this would increase the total flagging due to \ac{RFI} across all bands to \qty{15}{\percent}. The remainder of the mid band (as it appears in \ac{CASDA}) is the least affected by flagging due to other sources of \ac{RFI}. Three primary service allocations correspond to this band. The middle panel of Figure \ref{fig:EBL} shows that no more than \qty{2}{\mega\hertz} across \qty{75}{\percent} of the most affected channels is lost to flagging due to \ac{RFI}. The three most prevalent sources of flagging correspond to transmitters in the aeronautical radionavigation, radio location and earth exploration satellite primary service allocations. \ac{ASKAP}'s mid band is of particular importance as it contains the biggest protected \ac{RAS} allocation centered on the \qty{1420}{\mega\hertz} hydrogen line. In this allocation up to \qty{0.4}{\mega\hertz} is lost due to flagging; i.e \qty{1.5}{\percent} data in the hydrogen line protected \ac{RAS} allocation is flagged.

Whilst there are comparatively less data for the high band, the average flagging is \qty{11}{\percent}. The high band is the most complex with regards to the effects of \ac{RFI}. Fourteen services affect this band. The bottom panel of Figure \ref{fig:EBL} shows that the majority of data in the high band is lost to channels corresponding to an aeronautical radionavigation allocation (cyan) accounting for between \qtyrange{12}{25}{\mega\hertz} of average bandwidth loss. Other primary services affecting the high band are satellite (pink) and meteorological related, the former of which cumulatively accounts for a further \qty{10}{\mega\hertz} loss on average. There is currently insufficient data to estimate the losses in the narrower OH spectral line \ac{RAS} allocations. 

We compared the estimated losses to the measured \ac{RFI} monitoring occupancy, that is the average percentage of detected \ac{RFI} in a channel using the \ac{RFI} monitoring equipment since 2019. We find that in general, across the clean bands there is more agreement between \ac{ASKAP} and the more sensitive \ac{RFI} monitor. Both \ac{RFI} monitors find fixed/mobile  to be the most significant cause of interference in the low band. \ac{RFI} monitoring equipment does however detect more interference due to aeronautical mobile and radionavigation service related interference compared to \ac{ASKAP} flagging data. Similarly, in the mid band there is alignment between services and the more sensitive \ac{RFI} monitor. More high band observations are required to make reliable comparisons with \ac{RFI} monitoring equipment.

\subsection*{RFI trends}
Several other trends in the probabilities of flagging due to \ac{RFI} based on Figure \ref{fig:HOD} were also explored including the expected bandwidth loss per service year-on-year, differences in the probability of flagging due to \ac{RFI} in the day compared to night and the effects of ducting as compared to RFI monitoring equipment.

\subsubsection*{Year-on-year}
Following on from the previous section, the estimated bandwidth loss per service can be further separated by year. There is only enough data to make this comparison in \ac{ASKAP}'s clean low and mid bands, shown below for a subset of primary spectrum services.

\begin{figure}[h!]
  \includegraphics[width=\textwidth]{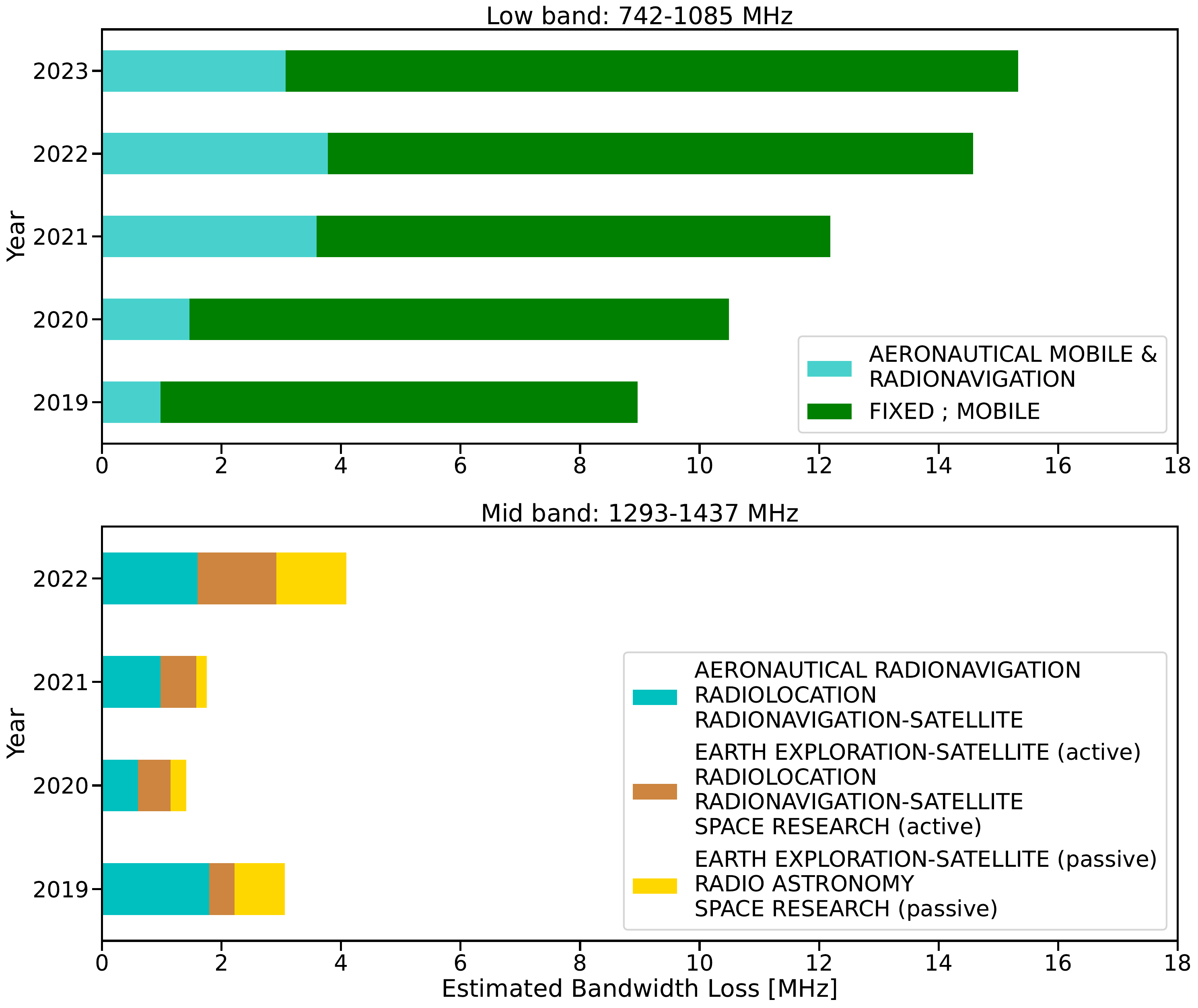}
  \caption{For the low band (top panel) and mid band (bottom panel) there is enough data to determine trends in the estimated bandwidth loss per service year-on-year. The increasing losses corresponding to fixed/mobile (green) is of concern in the low band. The (yellow) protected \ac{RAS} allocations in the mid band where in previous years there have been non-negligible losses require ongoing monitoring and investigation to asertain the source of the flagging.}
  \label{fig:year_on_year}
\end{figure}

Figure \ref{fig:year_on_year} again shows that the clean mid band is less affected by \ac{RFI} than the low band. In the low band we see again that the primary losses are due to fixed/mobile (green). The data also shows that there is a year-on-year increase in the amount of flagged data corresponding to fixed/mobile allocations. Interestingly, in the \ac{RFI} monitoring data this trend is not evident, which could suggest that the amount of interference year-on-year remains more or less constant, but since 2019 the \ac{ASKAP} pipeline has improved in identifying and flagging the interference. Figure \ref{fig:year_on_year} shows less flagging in the bands corresponding to aeronautical radionavigation service allocations (cyan) in 2019 and 2020, compared to after 2020 consistent with travel restrictions due to the COVID-19 pandemic. This pattern is not as clear in the mid band aeronautical radionavigation service allocations, nor do we see any obvious increases in the modest flagging in any other services. Of concern are the (yellow) protected \ac{RAS} allocations where in previous years there have been non-negligible losses. Further investigation is required using full resolution data from the \ac{SMART} project to determine if in some cases \texttt{CFLAG} is flagging the \qty{21}{\centi\meter} line itself or if this is due to RFI as a result of \acl{UEMR}.

\begin{figure*}[!h]
  \includegraphics[width=\textwidth]{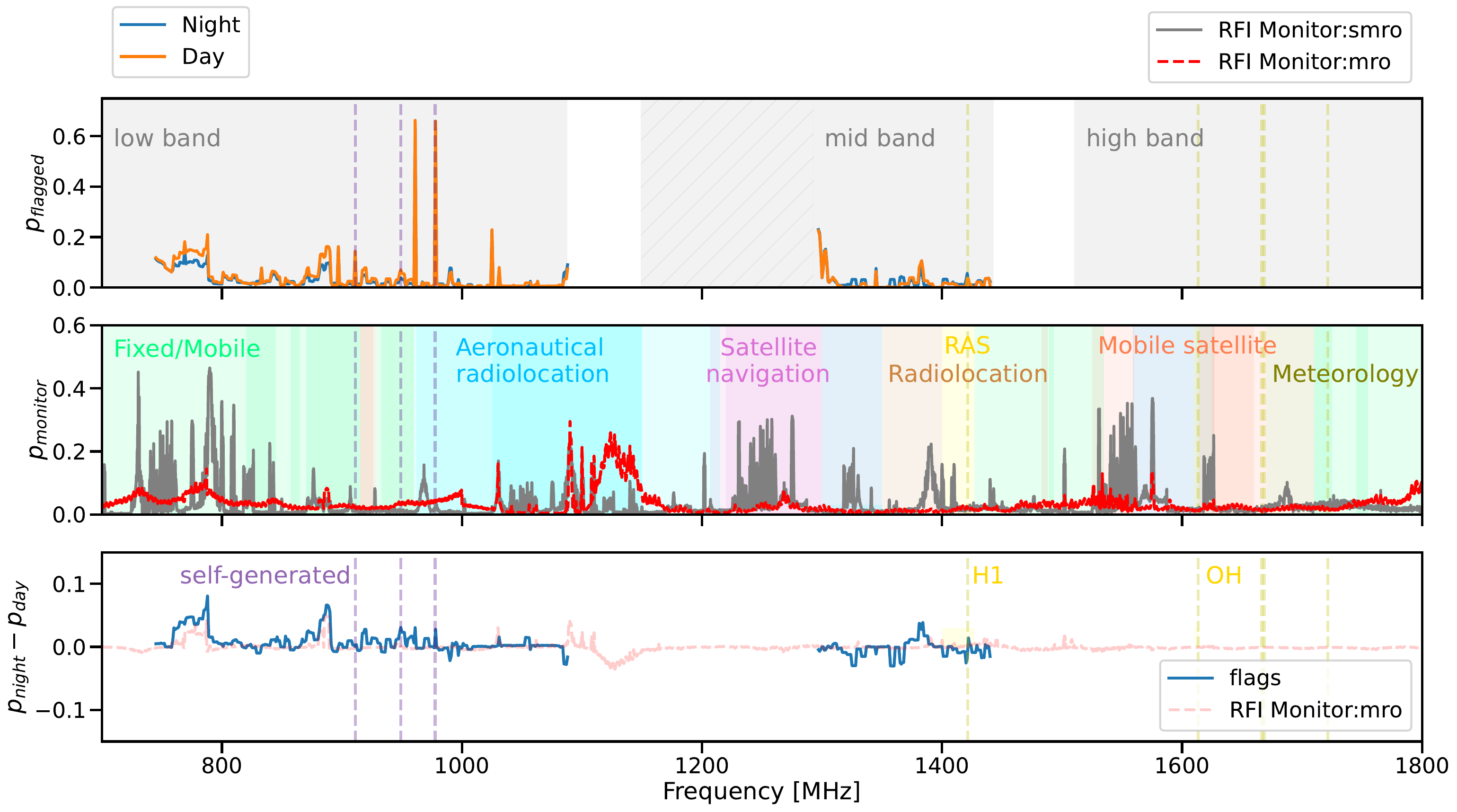}
  \caption{The top panel shows the mean probability of flagging, (from the blue vertical subplot in Figure \ref{fig:HOD}) comparing flagging in the day (orange) versus night (blue). The middle panel shows the probability of \ac{RFI} based on \ac{RFI} monitors. The bottom panel shows differences in flagging between night and day (blue) and \ac{RFI} monitor (opaque red dashed curve in the background). In the background of the top panel are \ac{ASKAP}'s low, mid and high bands (hatched area is discarded after observing). In the background of the second panel the broad ranges of several primary allocations affecting \ac{ASKAP}. Finally, the protected band (spectral lines) in yellow and the self-generated \ac{RFI} in purple are shown vertically across all panels.}
  \label{fig:day_vs_night}
\end{figure*}

\subsubsection*{Day vs Night}
Figure \ref{fig:HOD} shows some variation as a function of day and night. The red curve on the horizontal subplot is the easiest to discern by eye. Historically, it indicates that there have been more continuum science observations between 08:00 and 21:00 UTC (16:00 and 03:00 local time). This trend can be explained by maintenance on-site and beamforming occurring during the day. Going forward with full survey operations we expect this difference to be reduced. Note, \ac{ASKAP} does not currently support \acp{SSP} being observed only at night. Another obvious variation between day and night is in the main plot below \qty{800} {\mega\hertz} and at \qty{900} {\mega\hertz} corresponding to the fixed/mobile service which shows more flagging after sunset (19:00 local time 11:00 UTC).

To quantitatively determine the variations between day and night, Figure \ref{fig:day_vs_night} plots probabilities due to \ac{RFI} in the day (orange) and night (blue) as a function of frequency in the top panel; i.e. the blue vertical subplot in Figure \ref{fig:HOD} split into flagging in the day versus night. We have defined the daytime as 06:00 to 18:00 local time (inclusive) based on the average sunset and sunrise times across the year corresponding to 22:00 to 10:00 UTC in Figure \ref{fig:HOD}. This division also yields two equal \qty{12}{\hour} subsets of data with which to make the comparison. The middle panel in Figure \ref{fig:day_vs_night} shows the probability of \ac{RFI} based on two \ac{RFI} monitors at the observatory with different integrating times. The \textit{smro} (solid grey) \ac{RFI} monitor is more sensitive to transmission from mobile base stations and satellites, the \textit{mro} (dashed red) \ac{RFI} monitor gives better estimates on transmission from aeronatical radionavigation equipment, for example \ac{DME}. The bottom panel shows the difference between day and night in flagging (blue) and the \textit{mro} \ac{RFI} monitor (red). The ranges of the clean low, mid and high bands are shown in the background of the top panel. The ranges of several primary radio allocations affecting \ac{ASKAP} are also shown in the middle panel. The \ac{RAS} protected bands and spectral lines are shown in yellow. Finally, the candidate self-generated \ac{RFI} signals from Figure \ref{fig:self-generated} are overlayed vertically across all three panels in purple. 

The bottom panel in Figure \ref{fig:day_vs_night} shows variations in flagging between day and night (the solid blue curve) are within $\pm$\qty{10}{\percent}. This curve traces the \textit{mro} \ac{RFI} monitor. Positive values in the bottom panel indicate that the interference and flagging are worse at night and negative values indicate the interference and flagging are worse during the day. The most prominent examples of variations between day and night are in the low band, the fixed/mobile below \qty{800}{\mega\hertz} and at around \qty{870}{\mega\hertz} is more likely during the night. At first one might expect that \ac{RFI} due to cellular interference would decrease at night when people are generally asleep. However, recall that \ac{ASKAP} is located in the \ac{ARQZWA}, this means that \ac{RFI} from mobile base stations (see Figure \ref{fig:basestationMap_Balt}) is more likely to affect \ac{ASKAP} during ducting events. Indeed, upon comparison with \ac{RFI} monitoring ducting predictors since 2019, ducting is \qty{9}{\percent} more likely on average at night.

\subsubsection*{Ducting}
We have tried to identify the percentage of scheduling blocks affected by ducting based on `excess' flagging compared to the mean across all scheduling blocks. Table \ref{tab:excess_flagging} shows the percentage of scheduling blocks per band with probabilities greater than the mean, $1\sigma$ and $2\sigma$ above the mean. Across all data, day and night, and across all seasons we estimate an upper limit of \qty{12}{\percent} of the processed scheduling blocks affected by `excess' flagging potentially due to ducting. There are not yet any discernible seasonal trends in `excess' flagging though, across the year. \cite{Sokolowski2016TheObservatory} observed ducting in \qty{28}{\percent} of 131 nights in summer at the observatory at low frequencies ($\leq$\qty{300}{\mega\hertz}), with which we expect some correlation during periods of intense ducting at \ac{ASKAP}'s frequency range \citep{Indermuehle2017TheBETA}. 

\begin{table}[]
\caption{Percentage of scheduling blocks with excess flagging.}
\label{tab:excess_flagging}
\begin{tabular}{@{}lrr@{}}
\toprule
       & low band (\%) & mid band (\%) \\ \midrule
$p_{RFI}>\mu$     & 23       & 39       \\
$p_{RFI}>1\sigma$ & 11       & 12       \\
$p_{RFI}>2\sigma$ & 4        & 6        \\ \bottomrule
\end{tabular}
\end{table}

Predictors for \ac{RFI} ducting have been developed utilizing \ac{RFI} monitoring equipment data spanning the last five years. These predictors differentiate between four levels of ducting status: none, mild, moderate, and severe. We see mild ducting \qty{1.35}{\percent} of the time, moderate ducting \qty{4}{\percent} of the time and severe ducting \qty{3}{\percent} of the time. Broken up year-on-year, the percentage of time experiencing increased \ac{RFI} propagation is shown in Table \ref{tab:excess_propogation}.

\begin{table}[]
\caption{Percentage of year experiencing increased \ac{RFI} propagation.}
\label{tab:excess_propogation}
\begin{tabular}{@{}lrrr@{}}
\toprule
         & mild (\%)  & moderate (\%) & severe (\%) \\ \midrule
2023     & 0.05       & 4             & 5           \\
2022     & 3          & 5             & 2           \\
2021     & 0.62       & 1.40          & 0.47        \\
2020     & 1.35       & 2             & 1.18        \\
2019     & 1.92       & 7             & 5           \\ \bottomrule
\end{tabular}
\end{table}

\begin{figure}[!hb]
  \includegraphics[width=\textwidth]{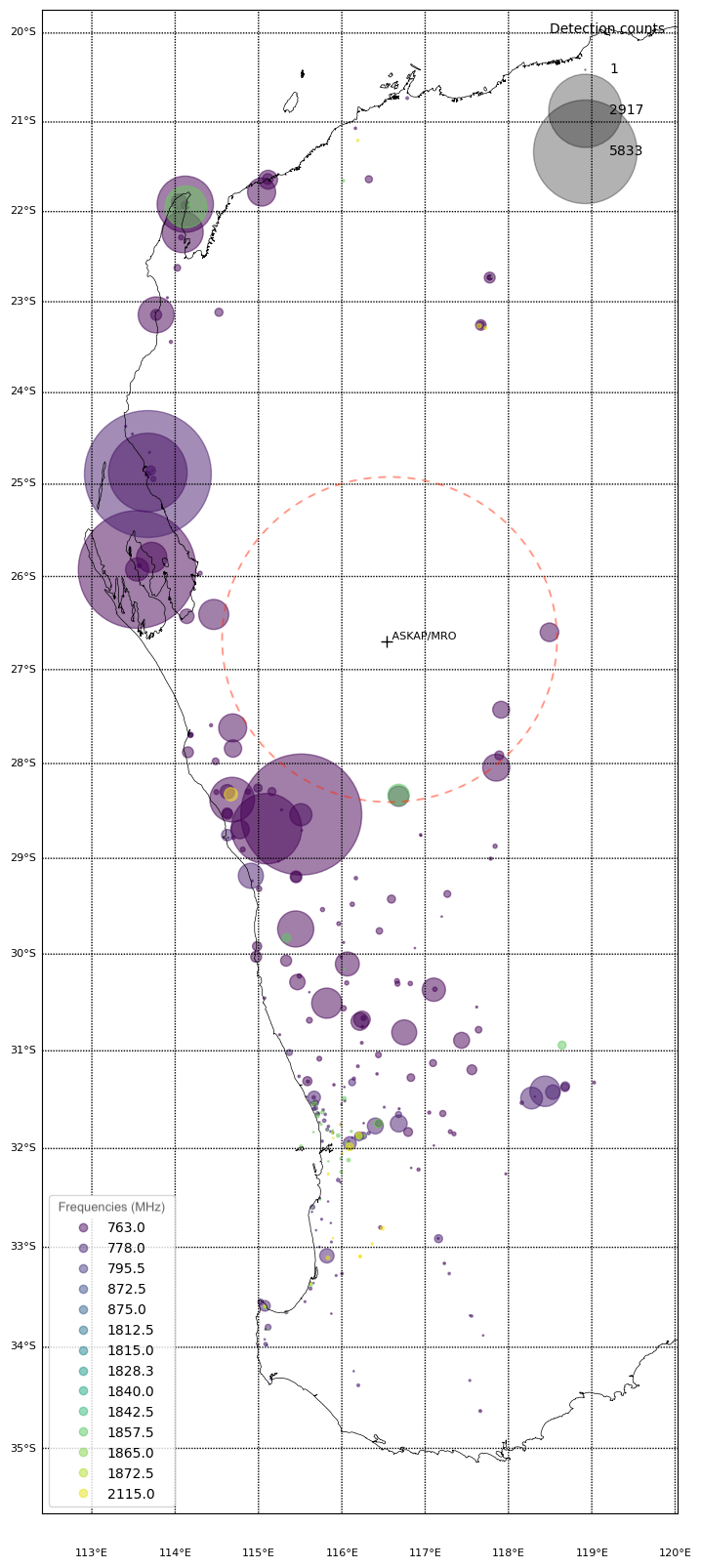}
  \caption{Map of mobile base station detections via \ac{ASKAP}'s \ac{RFI} monitoring equipment since 2017 showing the most frequently detected mobile base stations (the largest circles) are in the Mullewa/Geraldton (SW) or Carnarvon (NW) directions. The colour of the circles denotes the detected frequency. The map has been annotated to show the approximate area of the \ac{ARQZWA}.}
  \label{fig:basestationMap_Balt}
\end{figure}

Furthermore, Figure \ref{fig:basestationMap_Balt} shows the location of mobile base stations that have been detected at the observatory since 2017. Base stations with larger circles indicate more numerous and the colour  denotes the detected frequency. The majority of detected base stations correspond to the frequencies with increased flagging in the fixed/mobile service allocations below \qty{800}{\mega\hertz} and at around \qty{870}{\mega\hertz} at night shown in Figure \ref{fig:day_vs_night}. The most frequent detections of mobile base stations come from Mullewa/Geraldton (SW) and Carnarvon (NW) directions ($>$5000 detections). Detections from base stations this far (and further) away is only possible under circumstances in which there is severe ducting. Similarly, \ac{RFI} from maritime vessels are detected when there is ducting. Work is ongoing within the observatory to better characterise and detect ducting both historically and operationally, including work to automatically identify it in the data and reliably predict significant ducting events in advance. 

\section{Future Objectives}
\label{sec:future}
We will continue to process, aggregate and analyse archived \qty{1}{\mega\hertz} data from other \acp{SSP}. We plan to conduct regular follow-up \ac{RFI} surveys and are considering a mode in which we can sample the \ac{RFI} with full pointing coverage in \ac{ASKAP}'s fly's-eye mode. The first epoch of the dedicated \ac{RFI} survey is currently being carried out. The survey has been designed to complement our analysis of archived data. There are two primary motivations for carrying out an \ac{RFI} survey as part of the \ac{SMART} project. 

Firstly, frequency resolution: the data used to develop the \texttt{flagstats} library - and presented here within - comes primarily from archived data. The consequence of using archived data is that most of the data has a \qty{1}{\mega\hertz} frequency resolution. In contrast to narrowband \ac{RFI}, \qty{1}{\mega\hertz} is too wide to identify interference accurately. The dedicated survey visibilities (and flags) will help plan a mitigation strategy around the \ac{RFI} affecting \ac{ASKAP}. Furthermore, the survey will enable us to build out the software described in the previous section and routinely process flagged data from \ac{ASKAP} \qty{18.5}{kHz} resolution observations, thereby better understanding the origins of \ac{RFI} impinging on \ac{ASKAP} science.

Second, frequency coverage, from \qtyrange{1085}{1293}{\mega\hertz} and from \qtyrange{1437}{1510}{\mega\hertz} of the \ac{ASKAP} frequency range, is either entirely or frequently discarded after observing due to \ac{RFI}. As mentioned, \ac{ASKAP} \acp{SSP} generally observe in the three relatively clean bands to avoid \ac{RFI} caused by satellite-based radionavigation systems (GPS, Galileo, GLONASS, and BeiDou), aeronautical radio navigation (\acs{DME} and \acs{ADS-B}) and satellite-based radio communication systems (Iridium, Thuraya and Globalstar) \citep{Hotan2021AustralianDescription, Indermuehle2017TheBETA}. Collecting flagged data exclusively from \acp{SSP} results in large frequency gaps where we do not understand the \ac{RFI} environment or \ac{ASKAP}'s default flagging behaviour (see Figures \ref{fig:HOD} and \ref{fig:day_vs_night}).

We are currently working on extending the software presented herein to be compatible with observations in \ac{ASKAP}'s nominal full-band spectral-line resolution of \qty{18.5}{\kilo\hertz} (i.e. excepting zoom modes), starting with this first epoch of survey data, which will allow us to more accurately identify interferers and plan a mitigation strategy around them. We also plan to include flagged data as a function of pointing and baseline/antenna into our analysis. Full frequency survey data will be used to calibrate and infer the suitability of \ac{RFI} monitoring equipment. As part of \ac{ASKAP}'s holistic \ac{RFI} strategy we aim to more closely integrate the \texttt{flagstats} library and \ac{RFI} monitoring equipment, in particular, the full frequency telescope flagging data and high-time resolution \ac{RFI} monitor data 
(not presented here). Finally we plan to use the \ac{RFI} survey and flagged data monitoring system presented to determine and monitor the effects of \ac{RFI} mitigation. This includes but is not limited to, in the short-term spatial nulling of the self-generated interferers (Lourenço \& Chippendale in prep) presented in Section \ref{sec:self-generated}, to recover these affected frequencies.

\section{Conclusion}
\label{sec:conclusion}
\ac{RFI}-affected data in radio astronomy is typically thrown away in a process called flagging so as not to affect scientific conclusions. The \acl{SMART} project has generated an automated software processing pipeline that ingests flagged data upon completion of an observation. The software makes it easy to visualise flagging statistics due to \ac{RFI} as a function of frequency and time
~and \acp{SSP}. Flagging statistics per observation are also stored and mapped to other sources of data to compare and analyse populations of observations in novel ways. We have presented here the implementation of this software and are currently undertaking the first in a series of regular epochs of an \ac{RFI} survey with \ac{ASKAP}.  

We have shown the results from over \nhours of \qty{1}{\mega\hertz} averaged data, mapped features in flagged data to their respective radio allocation to estimate the impacts on \ac{ASKAP} science by other services in the radio spectrum and self-generated interferers. We have placed an upper estimate of \qty{3}{\percent} on the average data lost to \ac{RFI} in the clean science bands. The (archived and processed) mid band is the least affected by flagging, while the low band is most affected by mobile telecommunication base stations. 

The ongoing monitoring of all flagged data and regular full-frequency resolution \ac{RFI} surveys, presented herein, are essential in improving the overall quality of science conducted with \ac{ASKAP} amidst an increasingly dynamic and evolving radio spectrum.

\section*{Acknowledgments}
This scientific work uses data obtained from Inyarrimanha Ilgari Bundara, the \acs{CSIRO} \acl{MRO}. We acknowledge the Wajarri Yamaji People as the Traditional Owners and native title holders of the observatory site. \acs{CSIRO}'s \ac{ASKAP} radio telescope is part of the \acl{ATNF} (\href{https://ror.org/05qajvd42}{https://ror.org/05qajvd42}). Operation of \ac{ASKAP} is funded by the Australian Government with support from the National Collaborative Research Infrastructure Strategy. \ac{ASKAP} uses the resources of the Pawsey Supercomputing Research Centre. Establishment of \ac{ASKAP}, Inyarrimanha Ilgari Bundara, the \acs{CSIRO} \acl{MRO} and the Pawsey Supercomputing Research Centre are initiatives of the Australian Government, with support from the Government of Western Australia and the Science and Industry Endowment Fund. This paper includes archived data obtained through the \acl{CASDA} (\href{https://data.csiro.au}{https://data.csiro.au}).

\bibliography{references}

\begin{thebibliography}{}
\expandafter\ifx\csname natexlab\endcsname\relax\def\natexlab#1{#1}\fi

\bibitem[{Anderson {et~al.}(2021)Anderson, Heald, Eilek, Lenc, Gaensler, Rudnick, Van~Eck, O'Sullivan, Stil, Chippendale, Riseley, Carretti, West, Farnes, Harvey-Smith, McClure-Griffiths, Bock, Bunton, Koribalski, Tremblay, Voronkov, \& Warhurst}]{Anderson2021EarlyCluster}
Anderson, C.~S., Heald, G.~H., Eilek, J.~A., {et~al.} 2021, Publications of the Astronomical Society of Australia, doi:10.1017/pasa.2021.4

\bibitem[{{Australian Communications and Media Authority}(2021)}]{ACMA2021}
{Australian Communications and Media Authority}. 2021, {Australian Radiofrequency Spectrum Plan | ACMA}

\bibitem[{Baan(2019)}]{Baan2019ImplementingScience}
Baan, W.~A. 2019, Journal of Astronomical Instrumentation, 8, 1

\bibitem[{Black {et~al.}(2015)Black, Jeffs, Warnick, Hellbourg, \& Chippendale}]{Black2015Multi-tierTelescope}
Black, R.~A., Jeffs, B.~D., Warnick, K.~F., Hellbourg, G., \& Chippendale, A. 2015, in 2015 IEEE Signal Processing and Signal Processing Education Workshop, SP/SPE 2015, 261--266

\bibitem[{Brown {et~al.}(2014)Brown, Hampson, Roberts, Beresford, Bunton, Cheng, Chekkala, Kiraly, Neuhold, \& Jeganathan}]{Brown2014DesignSystem}
Brown, A.~J., Hampson, G.~A., Roberts, P., {et~al.} 2014, in Proceedings - 2014 International Conference on Electromagnetics in Advanced Applications, ICEAA 2014

\bibitem[{Burd {et~al.}(2018)Burd, Mannheim, M{\"{a}}rz, Ringholz, Kappes, \& Kadler}]{Burd2018DetectingAlgorithm}
Burd, P.~R., Mannheim, K., M{\"{a}}rz, T., {et~al.} 2018, Astronomische Nachrichten, 339, doi:10.1002/asna.201813505

\bibitem[{Chapman {et~al.}(2017)Chapman, Dempsey, Miller, Heywood, Pritchard, Sangster, Whiting, \& Dart}]{Chapman2017CASDA:Archive}
Chapman, J.~M., Dempsey, J., Miller, D., {et~al.} 2017, in Astronomical Society of the Pacific Conference Series, Vol. 512, Astronomical Data Analysis Software and Systems XXV, ed. N.~P.~F. Lorente, K.~Shortridge, \& R.~Wayth, 73

\bibitem[{Chippendale \& Hellbourg(2017)}]{Chippendale2017InterferenceTelescope}
Chippendale, A.~P., \& Hellbourg, G. 2017, in 2017 International Conference on Electromagnetics in Advanced Applications (ICEAA) (IEEE), 948--951

\bibitem[{{Committee on Radio Astronomy Frequencies}(1997)}]{CommitteeonRadioAstronomyFrequencies1997CRAFAstronomy}
{Committee on Radio Astronomy Frequencies}. 1997, {CRAF Handbook for Radio Astronomy} (Dwingeloo: CRAF Secretariat)

\bibitem[{Cornwell {et~al.}(2016)Cornwell, Humphreys, Lenc, Voronkov, Whiting, Mitchell, Ord, Collins, \& Guzman}]{Cornwell2016ASKAPASKAP-SW-0020}
Cornwell, T., Humphreys, B., Lenc, E., {et~al.} 2016, {ASKAP Science Processing, ASKAP-SW-0020}

\bibitem[{{CSIRO}(2022)}]{CSIRO2022CflagDocumentation}
{CSIRO}. 2022, {cflag (Flagging Utility) — ASKAP Central Processor documentation}

\bibitem[{{Di Vruno F.} {et~al.}(2023){Di Vruno F.}, {Winkel B.}, {Bassa C. G.}, {J{\'{o}}zsa G. I. G.}, {Brentjens M. A.}, {Jessner A.}, \& {Garrington S.}}]{DiVrunoF.2023UnintendedMHz}
{Di Vruno F.}, {Winkel B.}, {Bassa C. G.}, {et~al.} 2023, A{\&}A, 676, A75

\bibitem[{Duchesne {et~al.}(2023)Duchesne, Thomson, Lenc, E, Moss, McConnell, Wieringa, Whiting, Wang, Wang, Rose, Raja, Leung, K, Huynh, Hotan, Hodgson, \& Heald}]{Duchesne2023TheRACS-mid}
Duchesne, S.~W., Thomson, A. J.~M., Lenc, J.~P., {et~al.} 2023, Publications of the Astronomical Society of Australia, 40, doi:10.1017/pasa.2023.31

\bibitem[{Fridman \& Baan(2001)}]{Fridman2001RFIAstronomy}
Fridman, P.~A., \& Baan, W.~A. 2001, Astronomy and Astrophysics, doi:10.1051/0004-6361:20011166

\bibitem[{{Grigg D.} {et~al.}(2023){Grigg D.}, {Tingay S. J.}, {Sokolowski M.}, {Wayth R. B.}, {Indermuehle B.}, \& {Prabu S.}}]{GriggD.2023DetectionAnalogue}
{Grigg D.}, {Tingay S. J.}, {Sokolowski M.}, {et~al.} 2023, A{\&}A, 678, L6

\bibitem[{Guzman {et~al.}(2016)Guzman, Chapman, Marquarding, \& Whiting}]{Guzman2016StatusOperations}
Guzman, J.~C., Chapman, J., Marquarding, M., \& Whiting, M. 2016, in Software and Cyberinfrastructure for Astronomy IV, Vol. 9913 (SPIE), 991311

\bibitem[{Guzman \& Humphreys(2010)}]{Guzman2010TheArchitecture}
Guzman, J.~C., \& Humphreys, B. 2010, in Software and Cyberinfrastructure for Astronomy, Vol. 7740 (SPIE), 77401J

\bibitem[{Hale {et~al.}(2021)Hale, Mcconnell, Thomson, Lenc, Heald, Hotan, Leung, Moss, Murphy, Pritchard, Sadler, Stewart, \& Whiting}]{Hale2021TheRelease}
Hale, C.~L., Mcconnell, D., Thomson, A.~J., {et~al.} 2021, Publications of the Astronomical Society of Australia, 38, doi:10.1017/pasa.2021.47

\bibitem[{Hall \& Barclay(1989)}]{Hall1989Radiowave-propagation}
Hall, M. P.~M., \& Barclay, L.~W. 1989, NASA STI/Recon Technical Report A, 90, 45603

\bibitem[{Hellbourg(2016)}]{Hellbourg2016InterferenceInterferometry}
Hellbourg, G. 2016, in International Geoscience and Remote Sensing Symposium (IGARSS)

\bibitem[{Hellbourg {et~al.}(2017)Hellbourg, Bannister, \& Hotarp}]{Hellbourg2017SpatialArray}
Hellbourg, G., Bannister, K., \& Hotarp, A. 2017, in Proceedings of 2016 Radio Frequency Interference: Coexisting with Radio Frequency Interference, RFI 2016

\bibitem[{Hellbourg {et~al.}(2012)Hellbourg, Weber, Capdessus, \& Boonstra}]{Hellbourg2012ObliqueAstronomy}
Hellbourg, G., Weber, R., Capdessus, C., \& Boonstra, A.~J. 2012, in 2012 IEEE Statistical Signal Processing Workshop, SSP 2012

\bibitem[{Hotan {et~al.}(2021)Hotan, Bunton, Chippendale, Whiting, Tuthill, Moss, McConnell, Amy, Huynh, Allison, Anderson, Bannister, Bastholm, Beresford, Bock, Bolton, Chapman, Chow, Collier, Cooray, Cornwell, Diamond, Edwards, Feain, Franzen, George, Gupta, Hampson, Harvey-Smith, Hayman, Heywood, Jacka, Jackson, Jackson, Jeganathan, Johnston, Kesteven, Kleiner, Koribalski, Lee-Waddell, Lenc, Lensson, Mackay, Mahony, McClure-Griffiths, McConigley, Mirtschin, Ng, Norris, Pearce, Phillips, Pilawa, Raja, Reynolds, Roberts, Roxby, Sadler, Shields, Schinckel, Serra, Shaw, Sweetnam, Troup, Tzioumis, Voronkov, \& Westmeier}]{Hotan2021AustralianDescription}
Hotan, A.~W., Bunton, J.~D., Chippendale, A.~P., {et~al.} 2021, Publications of the Astronomical Society of Australia

\bibitem[{Huynh {et~al.}(2020)Huynh, Dempsey, Whiting, \& Ophel}]{Huynh2020TheArchive}
Huynh, M., Dempsey, J., Whiting, M.~T., \& Ophel, M. 2020, in Astronomical Society of the Pacific Conference Series, Vol. 522, Astronomical Data Analysis Software and Systems XXVII, ed. P.~Ballester, J.~Ibsen, M.~Solar, \& K.~Shortridge, 263

\bibitem[{Indermuehle {et~al.}(2017)Indermuehle, Harvey-Smith, Wilson, \& Chow}]{Indermuehle2017TheBETA}
Indermuehle, B.~T., Harvey-Smith, L., Wilson, C., \& Chow, K. 2017, Proceedings of 2016 Radio Frequency Interference: Coexisting with Radio Frequency Interference, RFI 2016, 43

\bibitem[{{ITU-R}(2016)}]{RR2016}
{ITU-R}. 2016, {Radio Regulations}

\bibitem[{{Kemball A.} \& {Wieringa Mark}(2000)}]{CASAMs}
{Kemball A.}, \& {Wieringa Mark}. 2000, {AIPS++ Memo 229, MeasurementSet definition V2.0}

\bibitem[{Kesteven(2010)}]{Kesteven2010OverviewSystems}
Kesteven, M. 2010, Proceedings of Science, 107

\bibitem[{McConnell {et~al.}(2020)McConnell, Hale, Lenc, Banfield, Heald, Hotan, Leung, Moss, Murphy, O'Brien, Pritchard, Raja, Sadler, Stewart, Thomson, Whiting, Allison, Amy, Anderson, Ball, Bannister, Bell, Bock, Bolton, Bunton, Chippendale, Collier, Cooray, Cornwell, Diamond, Edwards, Gupta, Hayman, Heywood, Jackson, Koribalski, Lee-Waddell, McClure-Griffiths, Ng, Norris, Phillips, Reynolds, Roxby, Schinckel, Shields, Tremblay, Tzioumis, Voronkov, \& Westmeier}]{McConnell2020TheResults}
McConnell, D., Hale, C.~L., Lenc, E., {et~al.} 2020, Publications of the Astronomical Society of Australia, 1

\bibitem[{Murphy {et~al.}(2013)Murphy, Chatterjee, Kaplan, Banyer, Bell, Bignall, Bower, Cameron, Coward, Cordes, Croft, Curran, Djorgovski, Farrell, Frail, Gaensler, Galloway, Gendre, Green, Hancock, Johnston, Kamble, Law, Lazio, Lo, MacQuart, Rea, Rebbapragada, Reynolds, Ryder, Schmidt, Soria, Stairs, Tingay, Torkelsson, Wagstaff, Walker, Wayth, \& Williams}]{Murphy2013VAST:Transients}
Murphy, T., Chatterjee, S., Kaplan, D.~L., {et~al.} 2013, {VAST: An ASKAP survey for variables and slow transients}, doi:10.1017/pasa.2012.006

\bibitem[{Murphy {et~al.}(2021)Murphy, Kaplan, Stewart, O'Brien, Lenc, Pintaldi, Pritchard, Dobie, Fox, Leung, An, Bell, Broderick, Chatterjee, Dai, D'Antonio, Doyle, Gaensler, Heald, Horesh, Jones, McConnell, Moss, Raja, Ramsay, Ryder, Sadler, Sivakoff, Wang, Wang, Wheatland, Whiting, Allison, Anderson, Ball, Bannister, Bock, Bolton, Bunton, Chekkala, Chippendale, Cooray, Gupta, Hayman, Jeganathan, Koribalski, Lee-Waddell, Mahony, Marvil, McClure-Griffiths, Mirtschin, Ng, Pearce, Phillips, \& Voronkov}]{Murphy2021TheSurvey}
Murphy, T., Kaplan, D.~L., Stewart, A.~J., {et~al.} 2021, {The ASKAP Variables and Slow Transients (VAST) Pilot Survey}, doi:10.1017/pasa.2021.44

\bibitem[{Norris {et~al.}(2011)Norris, Hopkins, Afonso, Brown, Condon, Dunne, Feain, Hollow, Jarvis, Johnston-Hollitt, Lenc, Middelberg, Padovani, Prandoni, Rudnick, Seymour, Umana, Andernach, Alexander, Appleton, Bacon, Banfield, Becker, Brown, Ciliegi, Jackson, Eales, Edge, Gaensler, Giovannini, Hales, Hancock, Huynh, Ibar, Ivison, Kennicutt, Kimball, Koekemoer, Koribalski, Lpez-Snchez, Mao, Murphy, Messias, Pimbblet, Raccanelli, Randall, Reiprich, Roseboom, Rttgering, Saikia, Sharp, Slee, Smail, Thompson, Urquhart, Wall, \& Zhao}]{Norris2011EMU:Universe}
Norris, R.~P., Hopkins, A.~M., Afonso, J., {et~al.} 2011, Publications of the Astronomical Society of Australia, 28, doi:10.1071/AS11021

\bibitem[{Offringa {et~al.}(2010)Offringa, de~Bruyn, Biehl, Zaroubi, Bernardi, \& Pandey}]{Offringa2010Post-correlationMethods}
Offringa, A.~R., de~Bruyn, A.~G., Biehl, M., {et~al.} 2010, Monthly Notices of the Royal Astronomical Society, 405, doi:10.1111/j.1365-2966.2010.16471.x

\bibitem[{Offringa {et~al.}(2012)Offringa, Van De~Gronde, \& Roerdink}]{Offringa2012ADetection}
Offringa, A.~R., Van De~Gronde, J.~J., \& Roerdink, J.~B. 2012, Astronomy and Astrophysics, 539, doi:10.1051/0004-6361/201118497

\bibitem[{Offringa {et~al.}(2015)Offringa, Wayth, Hurley-Walker, Kaplan, Barry, Beardsley, Bell, Bernardi, Bowman, Briggs, Callingham, Cappallo, Carroll, Deshpande, Dillon, Dwarakanath, Ewall-Wice, Feng, For, Gaensler, Greenhill, Hancock, Hazelton, Hewitt, Hindson, Jacobs, Johnston-Hollitt, Kapi{\'{n}}ska, Kim, Kittiwisit, Lenc, Line, Loeb, Lonsdale, McKinley, McWhirter, Mitchell, Morales, Morgan, Morgan, Neben, Oberoi, Ord, Paul, Pindor, Pober, Prabu, Procopio, Riding, Udaya~Shankar, Sethi, Srivani, Staveley-Smith, Subrahmanyan, Sullivan, Tegmark, Thyagarajan, Tingay, Trott, Webster, Williams, Williams, Wu, Wyithe, \& Zheng}]{Offringa2015TheMitigation}
Offringa, A.~R., Wayth, R.~B., Hurley-Walker, N., {et~al.} 2015, Publications of the Astronomical Society of Australia, 32, doi:10.1017/pasa.2015.7

\bibitem[{Series(2013)}]{Series2013TechniquesAstronomy}
Series, R. 2013, Rep. ITU-R RAITU-R, 1, 1

\bibitem[{Sihlangu {et~al.}(2021)Sihlangu, Oozeer, \& Bassett}]{Sihlangu2021Multi-dimensionalObservatories}
Sihlangu, I., Oozeer, N., \& Bassett, B.~A. 2021, Journal of Astronomical Telescopes, Instruments, and Systems, 8, 1

\bibitem[{Sokolowski {et~al.}(2016)Sokolowski, Wayth, \& Lewis}]{Sokolowski2016TheObservatory}
Sokolowski, M., Wayth, R.~B., \& Lewis, M. 2016, in Conference Proceedings - GEMCCON 2015: IEEE Global Electromagnetic Compatibility Conference

\bibitem[{Tingay {et~al.}(2020)Tingay, Sokolowski, Wayth, \& Ung}]{Tingay2020AObservatory}
Tingay, S.~J., Sokolowski, M., Wayth, R., \& Ung, D. 2020, Publications of the Astronomical Society of Australia, doi:10.1017/pasa.2020.32

\bibitem[{Tuthill {et~al.}(2012)Tuthill, Hampson, Bunton, Brown, Neuhold, Bateman, de~Souza, \& Joseph}]{Tuthill2012DevelopmentASKAP}
Tuthill, J., Hampson, G., Bunton, J., {et~al.} 2012, in 2012 International Conference on Electromagnetics in Advanced Applications (IEEE), 1067--1070

\bibitem[{Wilson {et~al.}(2016)Wilson, Chow, Harvey-Smith, Indermuehle, Sokolowski, \& Wayth}]{Wilson2016TheMeasurements}
Wilson, C., Chow, K., Harvey-Smith, L., {et~al.} 2016, in Proceedings of the 2016 18th International Conference on Electromagnetics in Advanced Applications, ICEAA 2016

\bibitem[{Wilson {et~al.}(2015)Wilson, Storey, \& Tzioumis}]{Wilson2015MeasuresAustralia}
Wilson, C., Storey, M., \& Tzioumis, T. 2015, in 2013 Asia-Pacific Symposium on Electromagnetic Compatibility, APEMC 2013

\bibitem[{Zhang {et~al.}(2021)Zhang, Xu, Wang, Jing, Liu, Zhu, \& Jiang}]{zhang2021radio}
Zhang, C.-P., Xu, J.-L., Wang, J., {et~al.} 2021, {Radio Frequency Interference Mitigation and Statistics in the Spectral Observations of FAST}

\bibitem[{Zheleva {et~al.}(2023)Zheleva, Anderson, Aksoy, Johnson, Affinnih, \& DePree}]{Zheleva2023RadioCoexistence}
Zheleva, M., Anderson, C.~R., Aksoy, M., {et~al.} 2023, IEEE Communications Magazine, doi:10.1109/MCOM.005.2200389

\end{thebibliography}

\end{document}